\documentclass[a4paper,11pt]{article}
\usepackage{epsfig}
\usepackage[utf8]{inputenc}
\usepackage{cite}
\usepackage{hyperref}
\usepackage{epsfig}
\usepackage{amsmath}
\usepackage{amsfonts}
\usepackage{graphicx}
\usepackage{cite}
\usepackage{color}
\usepackage[dvipsnames]{xcolor}
\usepackage{multirow}
\usepackage{amssymb}
\usepackage{caption}
\usepackage{subcaption}

\graphicspath{{./Images/}}

\def\be{\begin{equation}}
\def\ee{\end{equation}}

\begin{document}
\title{{\bf  \LARGE  Lensing and shadow of a black hole \\
surrounded by a heavy accretion disk}}
 \author{
{\large Pedro V. P. Cunha}$^{1}$,\,
{\large Nelson A. Eiró}$^{2}$,\,
{\large Carlos A. R. Herdeiro}$^{3}$ and
{\large José P. S. Lemos}$^{2}$
\\
\\
$^{1}${\small Max Planck for Gravitational Physics - Albert Einstein Institute, } \\ {\small  Am M\"{u}hlenberg 1, Potsdam 14476, Germany}
\\
\\
$^{2}${\small Centro de Astrof\'isica e Gravitaç\~ao - CENTRA, Departamento de F\'isica,}\\ { \small Instituto Superior T\'ecnico - IST, Universidade de Lisboa - UL,}\\ {\small Av. Rovisco Pais 1, 1049-001, Lisboa, Portugal}
\\
\\
$^{3}${\small Departamento de Matem\'atica da Universidade de Aveiro and } \\ {\small  Centre for Research and Development  in Mathematics and Applications (CIDMA),} \\ {\small    Campus de Santiago, 3810-183 Aveiro, Portugal}
}
\date{December 2019}
\maketitle

\begin{abstract}
We consider a static, axially symmetric spacetime describing the superposition of a Schwarzschild black hole (BH) with a thin and heavy accretion disk. The BH-disk configuration is a solution of the Einstein field equations within the Weyl class. The disk is sourced by a distributional energy-momentum tensor and it is located at the set of fixed points of the geometry's $\mathbb{Z}_2$ symmetry, i.e., at the equatorial plane. It can be interpreted as two streams of counter-rotating particles, yielding a total vanishing angular momentum. The phenomenology of the composed system depends on two parameters: the fraction of the total mass in the disk, $m$, and the location of the inner edge of the disk, $a$. We start by determining the sub-region of  the space of parameters wherein the solution is physical, by requiring the velocity of the disk particles to be sub-luminal and real. Then, we study the null geodesic flow by performing backwards ray-tracing under two scenarios. In the first scenario the composed system is illuminated by the disk and in the second scenario the composed system is illuminated by a far-away celestial sphere.
Both cases show that, as $m$ grows, the shadow becomes more prolate. Additionally, the first scenario makes clear that as $m$ grows, for fixed $a$, the geometrically thin disk appears optically enlarged,  i.e., thicker, when observed from the equatorial plane. This is to due to light rays that are bent towards the disk, when backwards ray traced. In the second scenario, these light rays can cross the disk (which is assumed to be transparent) and may oscillate up to a few times before reaching the far away celestial sphere.  Consequently, an almost equatorial observer sees different patches of the sky near the equatorial plane, as a chaotic ``mirage". Indeed, for neighbouring observation angles,  different sets of oscillating light rays can end up on distinct regions of the far away sky. As $m\rightarrow 0$ one recovers the standard test, i.e., negligible mass, disk appearance.
 \end{abstract}

\newpage

\tableofcontents


\section{Introduction}
\begin{figure}[h!]
\begin{center}
\vspace{0.1cm}
\includegraphics[width=0.49\textwidth]{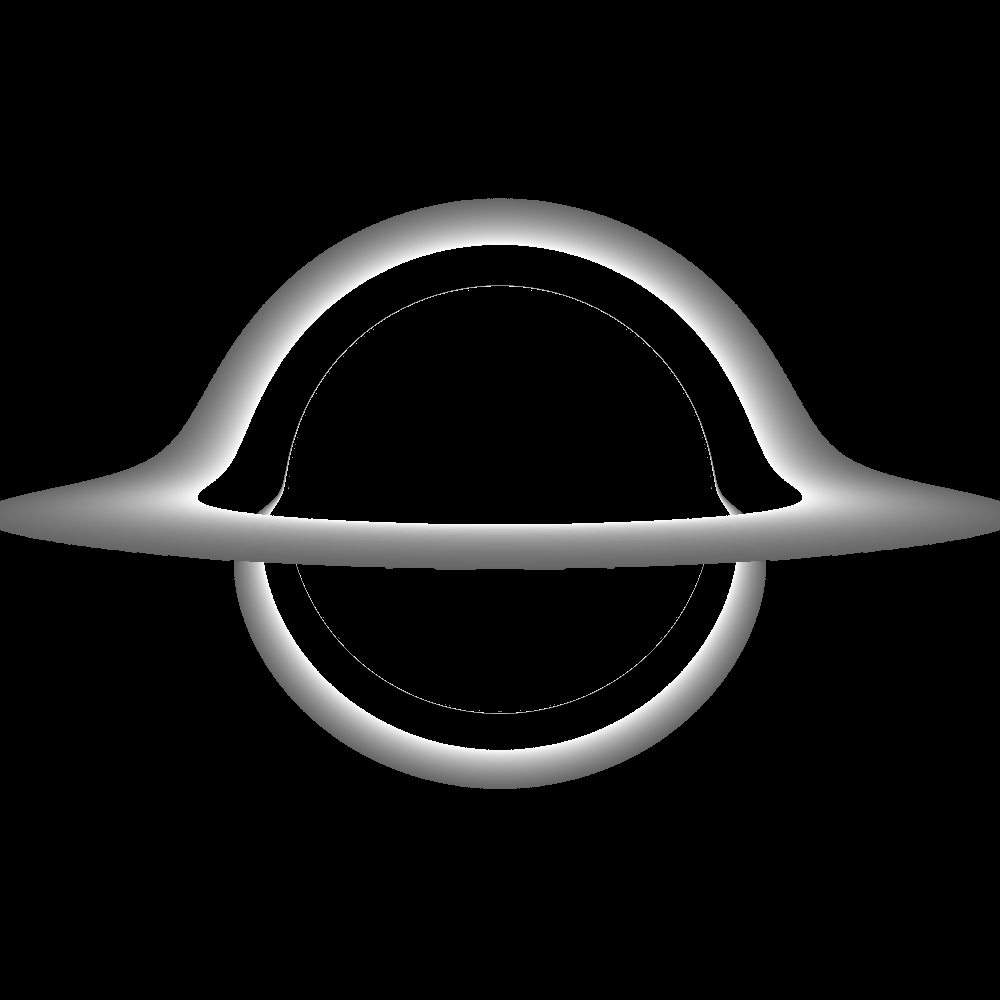}
\caption{\small A Schwarzschild BH surrounded by a thin, test, accretion disk. The disk is static. So there is no left-right asymmetry, unlike the classical image in~\cite{Luminet:1979nyg}. Radiation is sourced by the disk, which is on the equatorial plane ($\theta=90º$) and it is opaque. It has an inner edge at the innermost stable circular orbit (ISCO), a decaying luminosity profile, and an outer edge at certain radius. The observer is at $\theta=86º$ and at an areal radius of $20M$, where $M$ is the BH mass. The frontal part of the disk (with respect to the observer) is seen at the equator, whereas the part behind the BH is seen above and below the BH shadow, due to light  bending. The BH shadow is surrounded by a thin bright ring, corresponding to the spherical photon orbit.}
\label{fig0}
\end{center}
\end{figure}

Ground-breaking observations by the Event Horizon Telescope collaboration provided the first imaging of horizon scale structure, and in particular the lensing ring of an astrophysical black hole (BH)~\cite{Akiyama:2019cqa,Akiyama:2019eap}. Many astrophysical BHs are believed to be surrounded by accretion disks. Typically, the latter are modelled as having a minor effect on the spacetime geometry, and treated as test disks. The first computer image of a BH surrounded by an emitting accretion thin disk was obtained in 1979 by Luminet, under this assumption~\cite{Luminet:1979nyg}. For comparison purposes with the results in this paper, an image of a Schwarzschild BH surrounded by an emitting, thin and negligible mass accretion disk, produced from the setup in this work, is exhibited in Fig.~\ref{fig0}. It is nonetheless theoretically, and perhaps even astrophysically, interesting to inquire what would be the effect of a heavy accretion disk, which backreacts non-negligibly on the spacetime geometry. How does such a disk distort the BH shadow and the lensing of light by a BH? This paper will address this question.

Tackling such question finds, however, the immediate hurdle that the Einstein field equations are non-linear. No superposition principle holds, in general, allowing a simple, preferably analytic, construction of a spacetime describing a BH with a heavy accretion disk, even if the solutions describing these two ingredients were separately known. There are, however, two remarkable instances where a superposition principle holds in the fully non-linear theory of general relativity, describing strong gravity systems.

The first instance occurs in Einstein-Maxwell theory. An appropriate ansatz makes the full Einstein-Maxwell equations reduce to a single harmonic equation. Such a linear equation admits the superposition of different harmonic functions, with poles at different spacetime points. This multi-centre solution was found, historically, by Majumdar~\cite{Majumdar:1947eu} and Papapetrou~\cite{Papapetrou:1948jw}. 
In electro-vacuum, it was correctly interpreted by Hartle and Hawking~\cite{Hartle:1972ya} as the spacetime of many extremal Reissner-Nordstr\"om BHs, in equilibrium due to a no-force condition, resulting from the cancellation of their mutual gravitational attraction by their precisely equal mutual electric repulsion. This is an example of a complete linearisation of the full Einstein equations which finds many cousin examples in supergravity theories - see e.g.~\cite{Duff:1994an} -, being closely connected to supersymmetry~\cite{Gibbons:1982fy,Tod:1983pm}.
Due to the necessity of gauge charges, however, it has met little astrophysical interest, and no connection to the setup of an astrophysical BH surrounded by a heavy accretion disk can be envisaged from this construction.

The second instance occurs in vacuum. Weyl~\cite{Weyl:1917gp} first considered the problem of static, axially symmetric solutions of the vacuum Einstein field equations. In the appropriate gauge, the field equations for one of the two metric functions, $\phi$, yield a harmonic equation in an (auxiliary) flat Euclidean 3-space. This equation admits a superposition principle. The second function, however, is determined non-linearly by the first function, so that Weyl solutions retain the non-linearities of the full theory. Nonetheless, one can superimpose two (or more) Weyl solutions in an unambiguous fashion by considering the sum of the corresponding $\phi$ solutions. In this way, for instance, a superposition of two~\cite{Bach} (or more, if collinear,~\cite{1964NCim...33..331I}) Schwarzschild BHs can be constructed. The BHs interact and are actually sustained by conical singularities~\cite{Einstein:1936fp}. More relevant for our goal, one can superimpose a BH with a heavy accretion disk.

In 1994, Lemos and Letelier found a family of Weyl solutions describing a relativistic thin disk around a BH, with interesting physical properties~\cite{Lemos_Letelier}. The Lemos-Letelier disk, dubbed LL disk hereafter, has support on the equatorial plane only, wherein a non-zero energy-momentum tensor exists, defined as a Dirac delta function distribution. Moreover, the disk has an inner edge at some radial coordinate, making it possible, this way, to eliminate the necessity for unphysical matter in the energy-momentum tensor. Such unphysical matter was a necessity in previous Weyl disk-BH solutions, wherein the disk extended all the way until the event horizon~\cite{Morgan-Morgan}.

In this paper we shall use the composed Weyl solution describing the superposition of a Schwarzschild BH with the LL disk,
the BH+LL disk solution found in~\cite{Lemos_Letelier}, as a proxy to a realistic BH + heavy accretion disk system, to infer the impact of having a non-test accretion disk on the lensing and BH shadow - see e.g.~\cite{Perlick:2004tq,Cunha:2018acu} for reviews on lensing and BH shadows. To do so we shall perform backwards ray-tracing on this background using the code and setup we have developed previously - see e.g.~\cite{Cunha:2015yba,Cunha:2016bjh,Cunha_1605} and in particular~\cite{Cunha:2018gql,Cunha:2018cof} for previous applications to Weyl solutions. A second independent code was developed for this work, as reported below, with consistent results. As we shall see, a heavy accretion disk introduces qualitatively new features. In particular there is a deformation of the shadow shape, becoming more prolate as the disk contributes to a higher fraction of the total mass. Moreover, there is an optical enlargement of the thin disk, which appears thicker when observed from the equatorial plane, since light rays are bent towards the disk. Finally, when the disk is assumed to be transparent, light rays can oscillate a few times around the heavy disk and yield an observation of different patches of the sky, as a sort of chaotic mirage, near the (would be) image of the equatorial plane. An earlier study of timelike geodesics on this (and other) BH plus disk background(s) can be found in~\cite{Semerak:2012dw,Semerak:2012dx,Sukova:2013jxa,Witzany:2015yqa,Polcar:2019wfi}, where, in particular, chaotic behaviour was also described.

This paper is organised as follows. In section~\ref{section:Weyl} we briefly review Weyl solutions and present the Weyl description of the Schwarzschild BH, the LL disk solution, and the composed BH+LL disk system. In particular we analyse the parameter space determining the part wherein the physically most relevant solutions exist. In section~\ref{section:rt} we describe briefly the ray-tracing codes and the two physical setups that will be used in the next section. In section~\ref{section:Results}  we present our result, including a gallery of images obtained from the ray-tracing to assess the impact of a progressively more massive disk. The discussion of some photon trajectories is also presented to interpret the ray-tracing images. Closing remarks are presented in section~\ref{section:Conclusions}, where we exhibit the analogue of Fig.~\ref{fig0} for a composed BH+LL disk system, to visually illustrate in a didactic way the impact of a heavy accretion disk. An appendix discusses a particular feature in some ray-tracing images, that we dub ``earlobes". Natural units $G=1=c$ will be assumed throughout the paper.

\section{Weyl solutions}
\label{section:Weyl}

\subsection{Weyl ansatz and field equations}
The ansatz for a static and axisymmetric spacetime is given, in Weyl coordinates $(t,\rho,z,\varphi)$, by

\begin{equation}
  ds^2 = -e^{\phi(\rho,z)} dt^2 + e^{-\phi(\rho,z)}\left[e^{\nu(\rho,z)}(d\rho^2 + dz^2) + \rho^2 d\varphi^2\right] \ ,
  \label{eq:Weyl_general}
\end{equation}

\noindent where the metric functions $\phi(\rho,z)$ and $\nu(\rho,z)$ only depend on $\rho$ and $z$.
Inserting the ansatz~\eqref{eq:Weyl_general} into the vacuum Einstein's equations, $R_{\mu \nu} = 0$, yields a Laplace equation

\begin{subequations}

  \begin{equation}
  \Delta_{\mathbb{E}^3}\phi=0 \ ,
    \label{eq:Laplace}
  \end{equation}
where $\mathbb{E}^3$ is flat Euclidean 3-space in cylindrical coordinates and $\Delta$ is the Laplacian operator on this space. The function $\nu$ is then determined as a line integral from the derivatives of $\phi$:
  \begin{equation}
    \nu[\phi] =\frac{1}{2}\int{\rho \left[(\phi_{,\rho}^2 -\phi_{,z}^2 ) d\rho + 2\phi_{,\rho}\phi_{,z} dz \right]} \ .
    \label{eq:nu_general}
  \end{equation}
\end{subequations}
The linearity of Laplace's equation guarantees that a superposition of two or more solutions is still a solution. However, the same does not hold for $\nu[\phi]$. If $\phi=\phi_1+\phi_2$, then, from \eqref{eq:nu_general}:

\begin{equation}
  \nu[\phi_1+\phi_2] = \nu[\phi_1] + \nu[\phi_2] + \nu[\phi_1, \phi_2] \ ,
\end{equation}

\noindent where

\begin{equation}
  \begin{aligned}
    \nu[\phi_1,\phi_2] = \int \rho \big[({\phi_1}_{,\rho}{\phi_2}_{,\rho} -{\phi_1}_{,z}{\phi_2}_{,z} ) d\rho
                 + ({\phi_1}_{,\rho} {\phi_2}_{,z} + {\phi_1}_{,z} {\phi_2}_{,\rho}) dz \big] \ ,
  \end{aligned}
\end{equation}
encodes the impact of the non-linearities on $\nu$.

In the following we shall superpose a Schwarzschild BH and a thin massive LL disk using this formalism~\cite{Lemos_Letelier}.

\subsection{The Schwarzschild BH in Weyl coordinates}
\begin{figure}[h!]
\begin{center}
\includegraphics[width=0.49\textwidth]{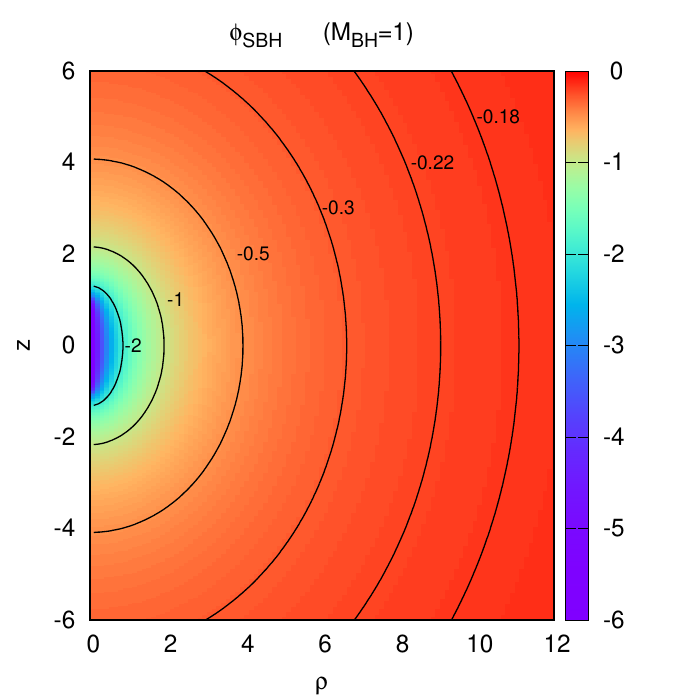}
\includegraphics[width=0.49\textwidth]{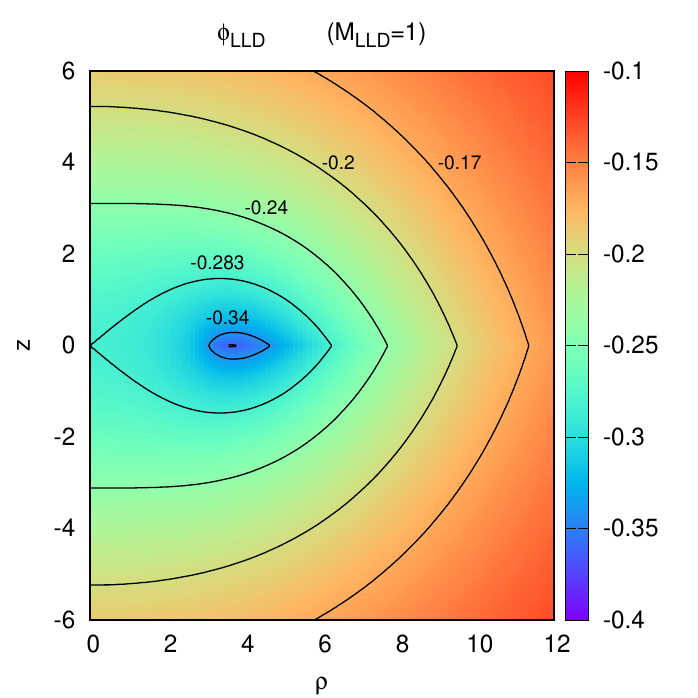}
\caption{\small The $\phi$ potential for the Schwarzschild BH (left panel) and LL disk (right panel) with $a=3.0$, in Weyl coordinates. Lines with $\phi=
{\rm constant}$ are represented in black. All quantities are normalised to the corresponding total mass.}
\label{fig1}
\end{center}
\end{figure}

The Schwarzschild BH in Weyl coordinates~\eqref{eq:Weyl_general} is described by the following $\phi=\phi_{\rm BH}$ function:
\begin{equation}
\phi_{\rm BH}=\log\left( \frac{d_+ +d_- -2M_{\rm BH}}{d_+ +d_- +2M_{\rm BH}}\right) \ ,
\label{phibh}
\end{equation}
where
\begin{equation}
d_\pm\equiv \sqrt{\rho^2+(z\pm M_{\rm BH})^2} \ .
\end{equation}
This $\phi$-potential is defined by a single parameter, the BH mass, $M_{\rm BH}$. This potential represents the Newtonian gravitational potential of an infinitesimally thin rod located at $\rho=0$ and $-M_{\rm BH}\leqslant z \leqslant M_{\rm BH}$. This becomes clear when plotting this potential, which is exhibited in Fig.~\ref{fig1} (left panel) in Weyl coordinates. The second metric function for the Schwarzschild BH in Weyl coordinates reads
\begin{equation}
\nu_{\rm BH}=\log\left[\frac{(d_++d_-)^2-4M_{\rm BH}^2}{4d_+ d_-} \right] \ .
\end{equation}

\subsection{The LL disk}
\subsubsection{The metric potentials of the LL disk}

The LL disk was obtained from an earlier disk solution, the Morgan-Morgan disk~\cite{Morgan-Morgan}. The latter possesses an outer edge, but not an inner edge. This makes it less physically relevant, as some of the particles that compose the disk
would extend all the way up to the location of the BH horizon and would possess tachyonic speeds~\cite{Lemos_Letelier}. The LL disk~\cite{Lemos_Letelier} is then found by inverting the Morgan-Morgan disk. This yields a static disk with an inner edge, made of two streams of counter-rotating particles \cite{LyndenBell1,LyndenBell2,Lemos_velocity}, with ``as many particles rotating to one side as to the other" \cite{Lemos_Letelier}. The result of this counter-rotation is a zero net angular momentum, which allows the existence of a static disk in equilibrium with the BH.

The Morgan-Morgan disk potential \cite{Morgan-Morgan} is most easily expressed in ellipsoidal coordinates $(\xi, \eta)$, where $-1 \leqslant \eta \leqslant 1$ and $0 \leqslant \xi < \infty$:
\begin{equation}
\phi_{\rm MMD} = - \frac { 2 M_{\rm MMD} } { a } \left\{ \cot^{-1}{\xi} + \frac { 1 } { 4 } \left[ \left( 3 \xi ^ { 2 } + 1 \right) \cot^{-1}{\xi} - 3 \xi \right] \times \left( 3 \eta ^ { 2 } - 1 \right)\right\} \ .
 \label{eq:MorganMorgan}
\end{equation}
The ellipsoidal coordinates are related to Weyl coordinates by:
 \begin{equation}
   \rho^2 = a^2(1+\xi^2)(1-\eta^2) \ , \qquad
   z = a \xi \eta \ .
    \label{eq:oblate}
 \end{equation}
Eq.~\eqref{eq:MorganMorgan} depends on two parameters: $M_{\rm MMD}$ is the mass of the disk and $a$ sets the outer edge of the disk, which is at  $\rho=a$.\\

 The inversion yielding the LL disk is a Kelvin transformation~\cite{Kelvin}, that maps harmonic functions into harmonic functions. This transformation acts on  the Weyl coordinates  as
\begin{equation}
 (\rho,z) \rightarrow \frac{a^2}{\rho^2 + z^2}(\rho,z) \ ,
 \label{eq:Kelvin}
\end{equation}
\noindent so that $\rho = a$ becomes now the inner edge of the LL disk.  The Kelvin transformation acts on the potential~\eqref{eq:MorganMorgan} as

\begin{equation}
\phi_{\rm LLD}(\rho,z) = \frac{a}{\sqrt{\rho^2 + z^2}}\phi_{\rm MMD}\left( \frac{a^2 \rho}{\rho^2 + z^2} , \frac{a^2 z}{\rho^2 + z^2} \right) \ .
\end{equation}
This yields:
\begin{equation}
  \begin{small}
    \begin{aligned}
    \phi_{\rm LLD}= & \frac{M_{\rm LLD} \sqrt{a^2+\chi-\rho ^2-z^2} \left(-3 a^2+3 \chi+\rho ^2+z^2\right)}{\sqrt{2} \pi  \left(\rho ^2+z^2\right)^2}  \\
                     & - \frac{2 M_{\rm LLD}\left[a^2 \left(2 z^2-\rho ^2\right)+2 \left(\rho ^2+z^2\right)^2\right]}{\pi  \left(\rho ^2+z^2\right)^{5/2}} \tan ^{-1}\left( \sqrt{\frac{2(\rho ^2+z^2)}{a^2+\chi-\rho ^2-z^2}}\right) \ ,
 \end{aligned}
\end{small}
\label{eq:Lamb_bar}
\end{equation}
where
\begin{equation}
\chi(\rho,z)\equiv \sqrt{a^4+2 a^2 (z^2-\rho^2 )+\left(\rho ^2+z^2\right)^2} \ .
\end{equation}

\noindent In~\eqref{eq:Lamb_bar} the relation $\cot^{-1}(\xi) = \tan^{-1}(1/\xi)$ ($\xi \geq 0$) was used.

Eq.~\eqref{eq:Lamb_bar} depends on two parameters: $M_{\rm LLD}$ is the mass of the LL disk and $a$ now sets the inner edge of the disk, which is at  $\rho=a$.
The relation between the masses $M_{\rm MMD}$ and $M_{\rm LLD}$ appearing in equations \eqref{eq:MorganMorgan} and \eqref{eq:Lamb_bar},  is~\cite{Lemos_Letelier}
\begin{equation}
M_{\rm LLD} =\frac{3}{4}\pi M_{\rm MMD} \ .
\end{equation}

The potential~\eqref{eq:Lamb_bar} is exhibited in Fig.~\ref{fig1} (right panel) in Weyl coordinates. The potential is sourced at the equatorial plane $z=0$. It is everywhere continuous, but (generically) not smooth at $z=0$. This allows us to define the energy-momentum tensor of the disk as a distribution. The $\nu$ metric function of the LL disk can be obtained either from~\eqref{eq:nu_general} or from the Kelvin transformation of the corresponding potential of the Morgan-Morgan disk. Its expression is, however, long and we shall not display it here.

\subsubsection{The energy-momentum tensor of the LL disk}

%

The $z$-derivatives of the LL disk metric, at the disk, are discontinuous. Expanding the metric immediately above (\textbf{+}) and below (\textbf{$-$}) the plane of the disk at $z = 0$ yields~\cite{Lemos_Letelier} (see \cite{arXiv_Huber} for a pedagogical treatment, see also
\cite{taub1980}):
\begin{equation}
 g_{\mu \nu} = g^0_{\mu \nu} + z\, {g^\pm_{\mu \nu}}_{,z}\big\rvert_{z=0} + \frac{1}{2} z^2 \, {g^\pm_{\mu \nu}}_{,z z} \big\rvert_{z=0} + \text{...} \ ,
 \label{eq:g_mu_nu_disk}
\end{equation}

\noindent where $g^0_{\mu \nu}$ is the value of the metric at $z = 0$. The discontinuities of the metric $z$-derivatives are denoted
\begin{equation}
 b_{\mu \nu} \equiv {g^+_{\mu \nu}}_{,z}\big\rvert_{z=0} - {g^-_{\mu \nu}}_{,z}\big\rvert_{z=0}\, .
 \label{eq:b}
\end{equation}
Then, the discontinuities of the Christoffel symbols are denoted as
\begin{equation}
  \begin{aligned}
    \left[\Gamma^\alpha_{\mu \nu}\right] \equiv  {\Gamma^{\alpha}}^+_{\mu \nu} - {\Gamma^{\alpha}}^-_{\mu \nu} = \frac{1}{2}(\delta^z_\mu b^\alpha_\nu + \delta^z_\nu b^\alpha_\mu - g^{\alpha z} b_{\mu \nu} ) \ .
  \end{aligned}
  \label{eq:Christoffel_z}
\end{equation}
The Riemann tensor is defined as a distribution on the disk, $R_{\alpha \beta \mu \nu}= \left[ R_{\alpha \beta \mu \nu} \right] \delta(z)$. One obtains for the Ricci tensor and scalar:
\begin{equation}
  \left[ R_{\beta \nu} \right] =  g^{\alpha \mu}\left[R_{\alpha \beta \mu \nu}\right]  = \frac{1}{2}\left( \delta^z_\beta b^z_\nu - g^{z z}b_{\nu \beta} + \delta^z_\nu b^z_\beta - \delta ^z_\beta \delta^z_\nu b_\alpha^\alpha \right) \ .
  \label{eq:RicciT_z}
\end{equation}
\begin{equation}
  \left[R\right] = g^{\beta \nu}\left[R_{\beta \nu}\right]= b^{z z} - g^{z z} b_\alpha^\alpha \ .
  \label{eq:RicciS_z}
\end{equation}
Thus, the distributed energy-momentum tensor $\left[ T^\mu_\nu \right]$ of the disk is obtained as
\begin{equation}
 8\pi \left[T^\mu_\nu\right]=  \left[R^\mu_\nu\right] - \frac{1}{2}\delta^\mu_\nu\left[R\right]  \ .
 \label{eq:dEE}
\end{equation}
The energy-momentum tensor, $T^\mu_\nu = \left[T^\mu_\nu\right]\delta(z)$, has the following non-zero components, in terms of the potentials $\phi,\nu$ and derivatives defined in the ($+$) branch:

\begin{subequations}
 \begin{equation}
   \epsilon = -T^t_t =  e^{\phi - \nu}(2 - \rho \phi_{,\rho} )\phi_{,z}\,\,\delta(z) \ ,
 \end{equation}
 \begin{equation}
   p_{\varphi \varphi} = T^\varphi_\varphi = e^{\phi- \nu}\rho \phi_{,\rho}\phi_{,z}\,\,\delta(z) \ ,
 \end{equation}
 \begin{equation}
   T^\rho_\rho = T^z_z = 0\ ,
 \end{equation}
 \label{eq:tensorT}
\end{subequations}

\noindent where $\epsilon$ is the energy density, $p_{\varphi \varphi}$ is the pressure density \cite{LyndenBell2}, and the factor of $8\pi$ is absorbed into the definition of $\left[T^\mu_\nu\right]$.

\subsection{The composed system: BH+LL disk }
Due to the linearity of the $\phi$ equation~\eqref{eq:Laplace}, the composed BH-disk system has a $\phi$ potential given by
\begin{equation}
\phi_{_{\rm BH+LLD}}=\phi_{\rm BH} +\phi_{\rm LLD} \ ,
\end{equation}
where the two pieces are given by~\eqref{phibh} and~\eqref{eq:Lamb_bar}, respectively. The $\nu$ potential can be obtained by integrating~\eqref{eq:nu_general}, which can always be done numerically.\\

The composed system is described by three parameters: the BH and disk masses, $M_{\rm BH}, M_{\rm LLD}$ and the $\rho$ coordinate of the inner edge of the disk, $a$. We can interpret the total mass as setting an overall scale. Moreover, we normalise the total mass to unity, i.e., we use units where the total mass sets the scale:
\begin{equation}
M_{\rm LLD}+M_{\rm BH}=1 \ .
\end{equation}
Thus, the phenomenology will depend on, say, the fraction of the total mass in the disk:
\begin{equation}
m\equiv \frac{M_{\rm LLD}}{M_{\rm LLD}+M_{\rm BH}} \ , \qquad m\in [0,1] \ .
\end{equation}

\begin{figure}[h!]
\begin{center}
\includegraphics[width=0.49\textwidth]{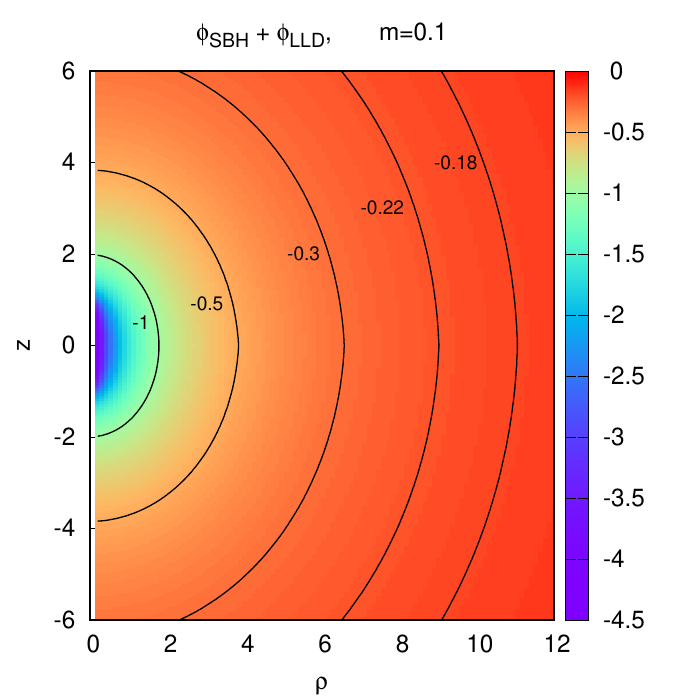}
\includegraphics[width=0.49\textwidth]{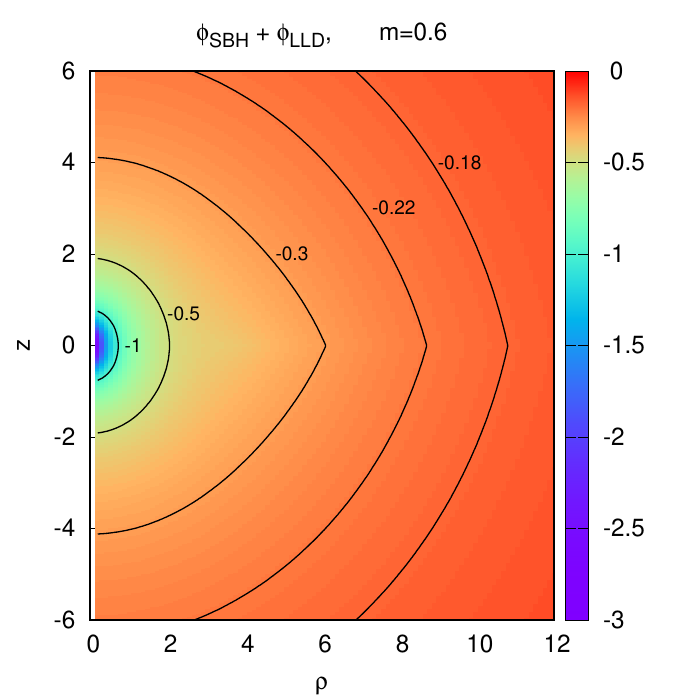}
\caption{\small The $\phi$ potential for the composed Schwarzschild BH plus LL disk system, in Weyl coordinates. (Left panel) $m=0.1$. (Right panel) $m=0.6$. For both cases $a=3.0$. As $m$ increases, corresponding to a disk with a larger fraction of the total mass, the potential $\phi_{\rm BH+LLD}$ becomes less smooth on the equator and the source on the $z$ axis decreases in size.}
\label{fig2}
\end{center}
\end{figure}

In Fig.~\ref{fig2} we plot the potential $\phi_{\rm BH+LLD}$ for two different mass ratios of the composed system, for the same value of $a$.

\subsection{The parameters space of astrophysical solutions}
As mentioned before, solutions describing Morgan-Morgan disks around a BH \textit{always} possess some particles with tachyonic speeds. This pathology can be cured by replacing the latter disks by LL disks; but since the inner edge of the disk is a free parameter, there are also solutions with this pathology within the LL class. Thus, one has to study what are the solutions describing the composed system of a BH with a LL disk that are physically sound.

\begin{figure}[h!]
   \centering
   \includegraphics[width=0.7\linewidth]{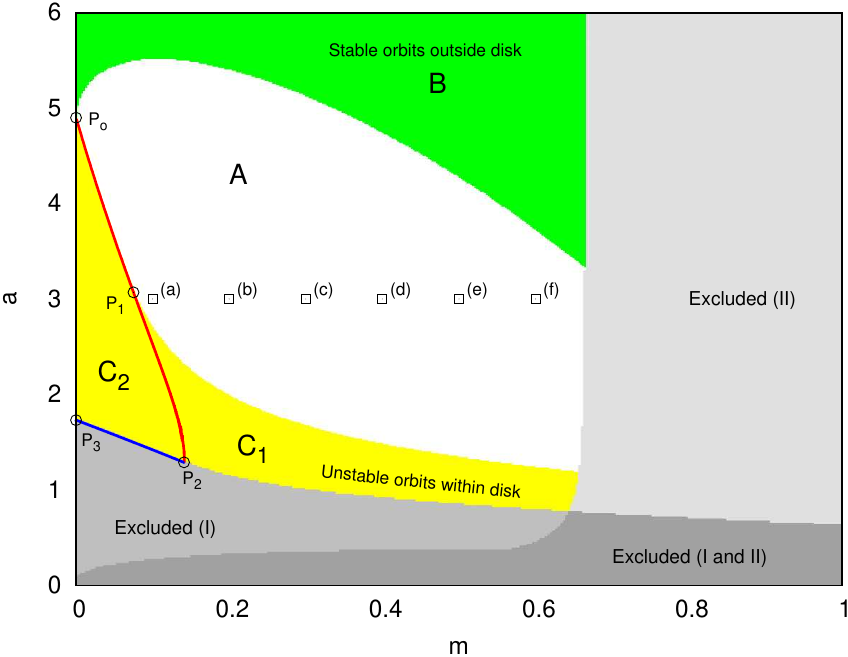}
   \caption{\small Assessment of the physical viability of the composed BH-disk solutions in the ($m$,$a$) domain. The regions in grey, classified as I or II, are physically excluded on the grounds of the disk having tachyonic particles, particles with complex speed or both (for different sets of particles). The non-excluded region is divided into subregions A, B and C  - see text for details, see also Fig.~\ref{fig:RegionsDrawing}.}
   \label{fig:AllowedRegions}
\end{figure}

The velocity of the disk particles is given by \cite{Lemos_Letelier,Lemos_velocity}
\begin{equation}
 V^2 = \frac{p_{\varphi \varphi}}{\epsilon} \ ,
 \label{eq:velocity}
\end{equation}

\noindent where $p_{\varphi \varphi}$ and $\epsilon$ are defined in \eqref{eq:tensorT}. A compact expression is provided by:
\[V^2=\frac{\kappa}{1-\kappa},\qquad\quad \kappa\,\equiv\, \frac{m}{\rho}\left(1-\frac{3}{2}\frac{a^2}{\rho^2}\right) + \frac{1-m}{\sqrt{\rho^2 + \left[1-m\right]^2}}.\]
A physical disk would necessarily satisfy ($0 \leqslant V^2 \leqslant 1$) for all $\rho\geqslant a$. If a spacetime obeys either of the two following conditions it must be excluded, on physical grounds:
\begin{itemize}
\item  (Exclusion type I):  $V^2>1$ for some $\rho\geqslant a$. The disk contains particles with tachyonic speeds;\\
\item  (Exclusion type II):  $V^2<0$ for some $\rho\geqslant a$. The disk contains particles with non-real speeds;\\
\end{itemize}
The exclusion conditions I and II are independent. For a subset of backgrounds, both can be satisfied, each at a different open set of points in the disk.  Using these conditions, we can find numerically which composed BH-disk systems are excluded in the space of parameters $\{m,a\}$ -  Fig.~\ref{fig:AllowedRegions}. The three regions in grey in Fig.~\ref{fig:AllowedRegions} are all unphysical, either satisfying condition I, II or both.

\begin{figure}[ht]
   \centering
   \includegraphics[width=0.6\linewidth]{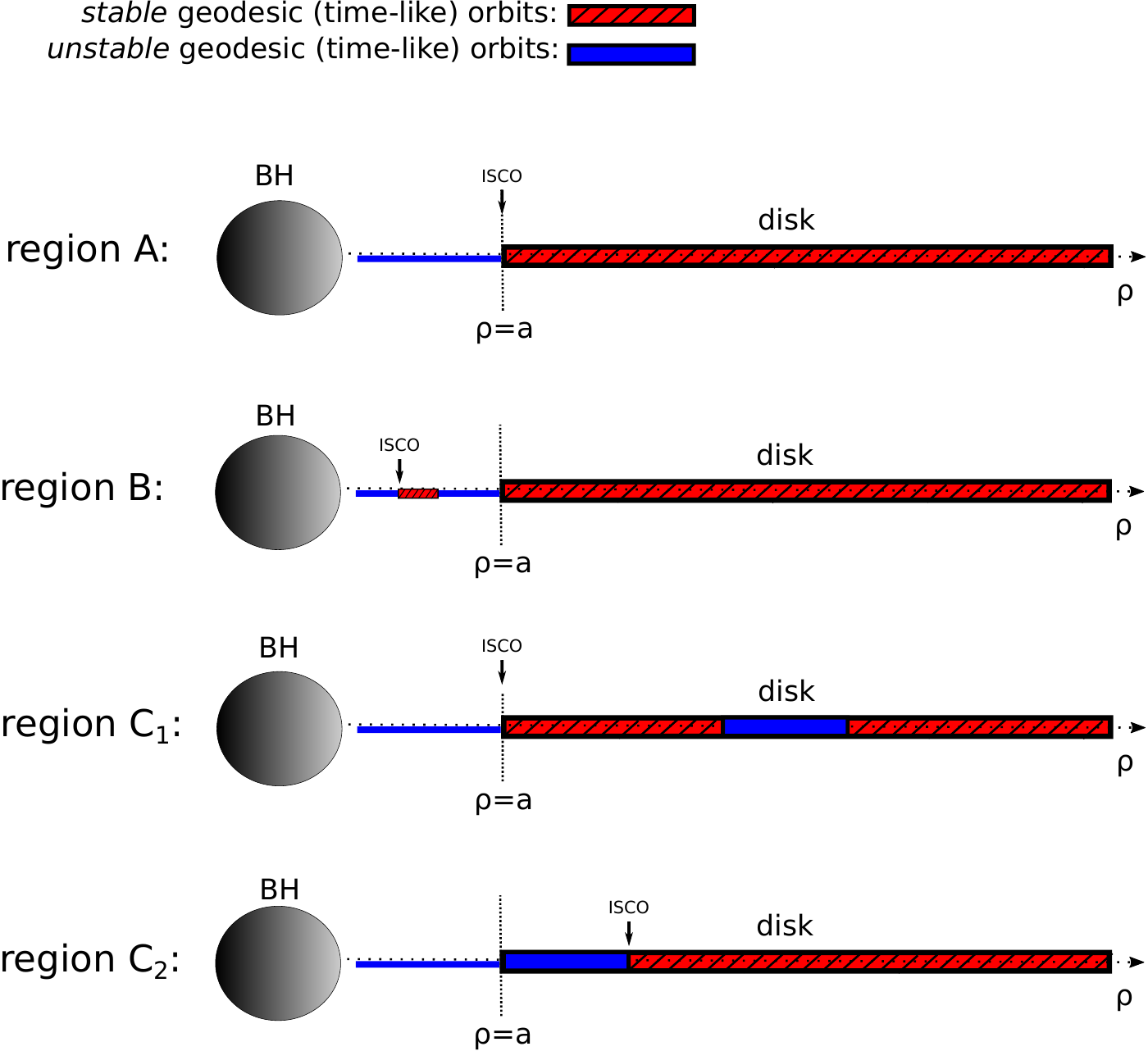}
   \caption{\small Schematic classification of the subregions in the non-excluded region in Fig.~\ref{fig:AllowedRegions}, according to the stability of circular timelike geodesics on the disk.}
   \label{fig:RegionsDrawing}
\end{figure}

Let us now turn to the non-excluded region. To perform a finer assessment of the physical properties of the solutions in this region we consider the stability of the circular geodesics in the disk region. First, we observe that the composed BH-disk system has, on the equatorial plane, an innermost stable circular orbit (ISCO). Recall that the ISCO for the Schwarzschild BH is located at an areal radial coordinate $r = 6\,M_{\rm BH}$, which in Weyl coordinates corresponds to $\rho=2\sqrt{6} M_{\rm BH}\simeq 4.9 M_{\rm BH}$. This is point $P_0$ in Fig~\ref{fig:AllowedRegions}. Then, one can divide the non-excluded solution domain into three sub-regions according to the following criteria, examining the stability of equatorial circular timelike geodesics, see Fig.~\ref{fig:AllowedRegions}:
\begin{itemize}
  \item In region $A$, the ISCO coincides with the inner edge of the disk (at $\rho_{_{\rm ISCO}}=a$). All equatorial circular orbits on the disk region are stable and all equatorial circular orbits outside the disk region are unstable;\\

  \item In region $B$, the ISCO is situated at  $\rho_{_{\rm ISCO}}<a$. All equatorial circular orbits on the disk region ($\rho\geqslant a$) are stable and some equatorial circular orbits outside the disk region are also stable (with $\rho_{_{\rm ISCO}}<\rho<a$);\\

  \item In region $C$ there are some unstable equatorial circular orbits inside the disk region. Region $C$ can be further subdivided into region $C_1$ (for which $\rho_{_{\rm ISCO}}=a$)  and into region $C_2$ (for which $a<\rho_{_{\rm ISCO}}$). In both cases all equatorial circular orbits outside the disk are unstable.
 \end{itemize}
For the sake of clarity, Fig.~\ref{fig:RegionsDrawing} contains an illustration of typical stability profiles within each subregion $A,B,C$.

The boundary between these regions merits some further discussion. Configuration $P_0$ in Fig.~\ref{fig:AllowedRegions} corresponds to the familiar case of a Schwarzschild BH with a massless disk ($m=0$), where its edge coincides with the ISCO $(\rho_{_\textrm{ISCO}}=a)$.
The red line in Fig.~\ref{fig:AllowedRegions}, starting from $P_0$, corresponds to configurations that possess a disk edge with marginally stable equatorial circular orbits. Between $P_0$ and $P_1$, this line separates subregions $A$ and $C_2$. Point $P_1$ is a bifurcation point. The remaining red line, between $P_1$ and $P_2$,  separates subregions $C_1$ and $C_2$; but from $P_1$ another boundary emerges, that between subregions $A$ and $C_1$, which terminates on the boundary of the excluded region II.
Point $P_2$ is at the boundary of the excluded region I. At this point the disk edge allows for circular photon orbits. Thus, the disk is on the verge of admitting tachyonic speeds (excluded region I). The blue curve starting from $P_2$ to $P_3$ corresponds to configurations wherein the disk edge is a circular photon orbit. The end point $P_3$, is a Schwarzschild BH with a massless disk ($m=0$) with an edge at its unstable circular photon orbit. This occurs  at the (areal) radial coordinate $r = 3\,M_{\rm BH}$, corresponding, in Weyl coordinates, to $\rho=\sqrt{3}\,M_{\rm BH}\simeq 1.7\,M_{\rm BH}$.

Although subregions $B$ and $C$ are not necessarily unphysical, we shall focus our analysis below on illustrative examples in region $A$, as accretion disk models with their edge at the ISCO are commonly used, e.g. the Novikov-Thorne model.  Thus, in our numerical imaging below, we will consider a sequence of solutions in region A, which are highlighted in Fig.~\ref{fig:AllowedRegions} as the sequence of points $(a) \,\to\, (f)$. These solutions have $a=3$ and $m=0.1;0.2;0.3;0.4;0.5;0.6$. We found this sequence of solutions representative of the generic features observed in region $A$, as $m$ is increased.

\section{Ray-tracing}
\label{section:rt}
\begin{figure}[h!]
\begin{center}
\includegraphics[width=0.3\textwidth]{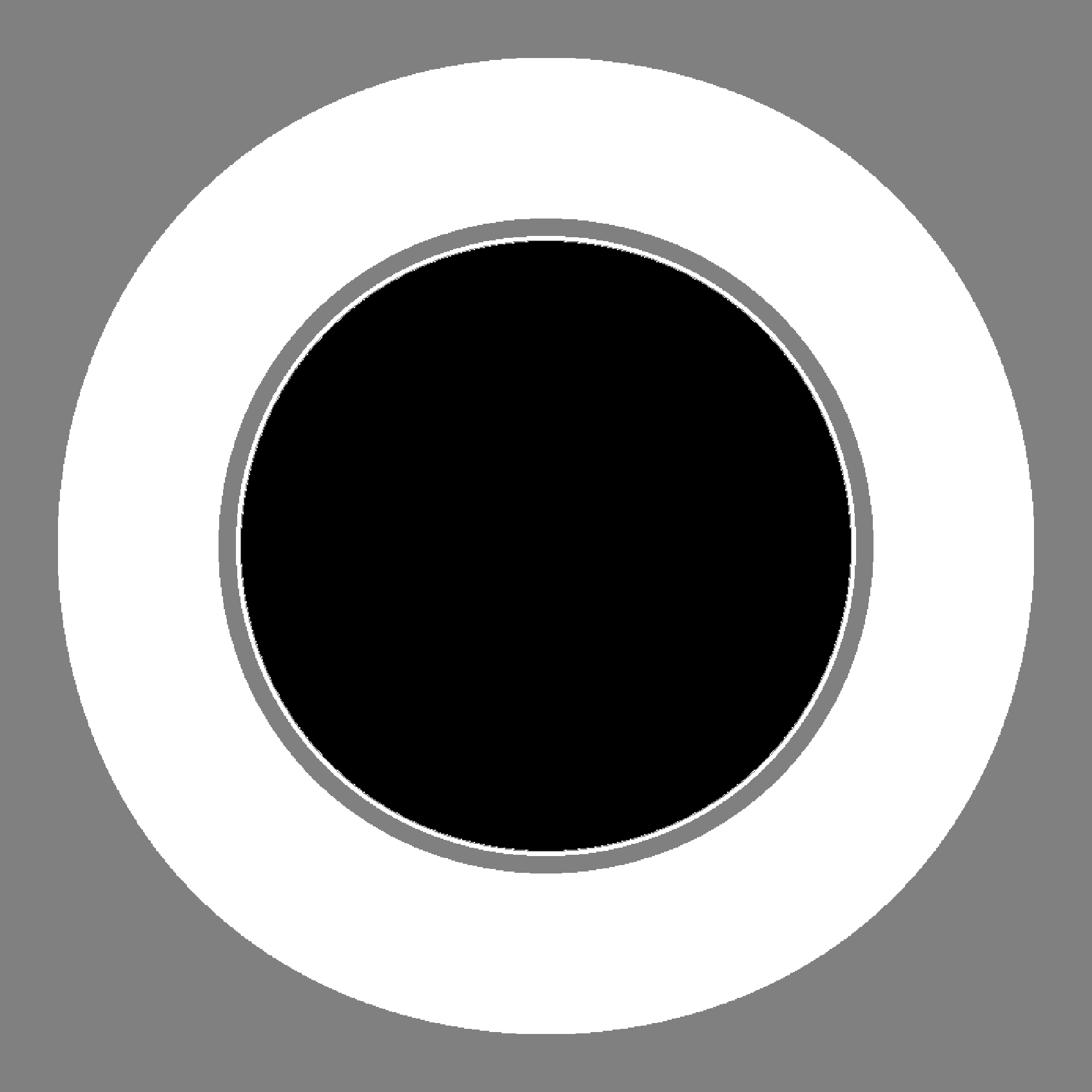}
\includegraphics[width=0.3\textwidth]{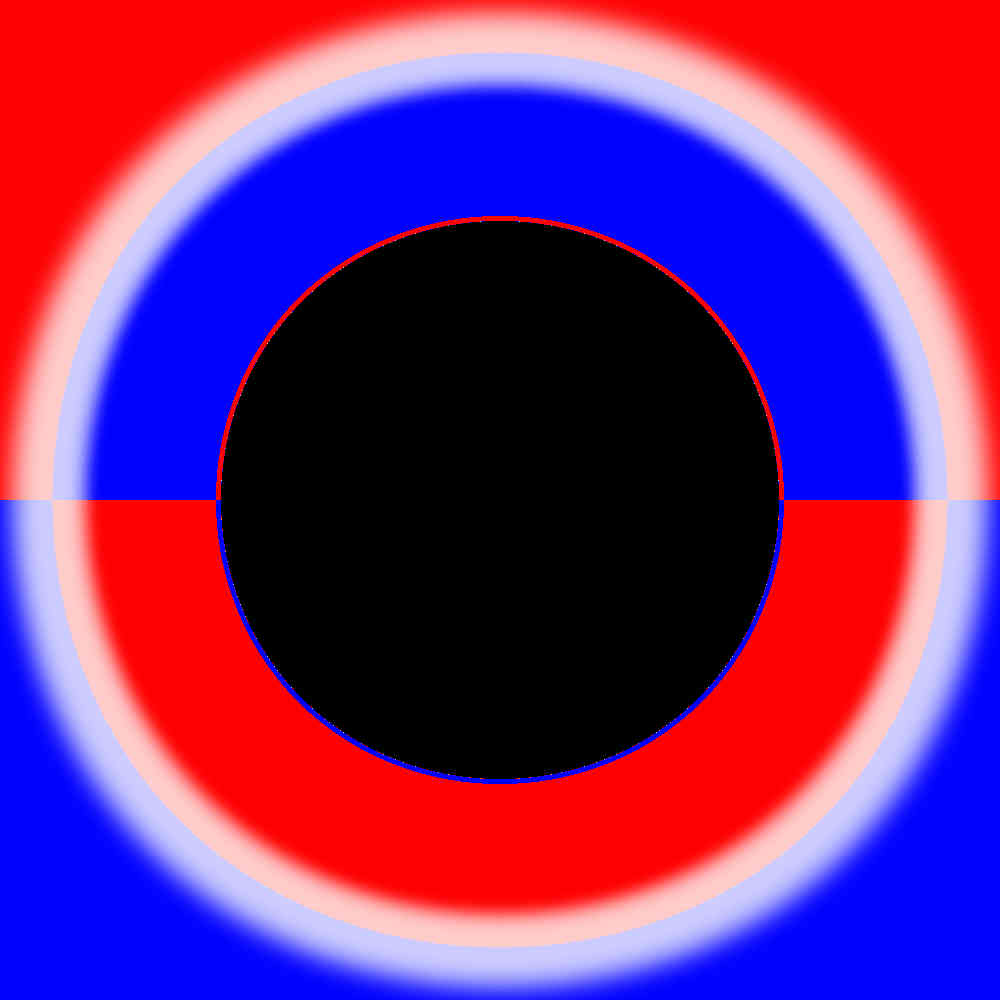}\hspace{0.05cm}
\caption{\small  Observation of a Schwarzschild BH surrounded a test accretion disk with inner edge at the ISCO and no outer edge, under the two scenarios described in Section~\ref{section:rt}.  (Left panel) Emission from the disk. (Right panel) Emission from a ``far away" two colours celestial sphere.}
\label{fig:scenarios}
\end{center}
\end{figure}

Two independent codes were used to image the shadow and lensing of the composed BH+LL disk system. The first code was the same used in e.g.~\cite{Cunha:2015yba,Cunha:2016bjh,Cunha_1605,Cunha:2018gql,Cunha:2018cof}. The second one was a code written in \textsc{python3}, and consists of a Runge-Kutta-Fehlberg algorithm \cite{iserles_2008} that numerically integrates the null geodesic equations $  \ddot{x}^\mu + \Gamma^\nu_{\alpha \beta} \dot{x}^\alpha \dot{x}^\beta = 0$. This numerical integration represents the propagation of photons from the observer backwards towards the source or the BH (backwards ray-tracing). We consider two scenarios:
\begin{description}
\item[{\bf I)}] In the first scenario the disk is the radiation source. The integration stops when the photon reaches either the BH, or the disk, or numerical infinity. Thus, the disk is opaque. Photons that end up (via backwards ray-tracing) on the disk/numerical infinity/BH are shown as white/grey/black pixels in the image. For this scenario, the visualisation of a Schwarzschild BH surrounded by a test accretion disk, with inner edge at the ISCO, is shown in Fig.~\ref{fig:scenarios} (left panel).
\item[{\bf II)}] In the second scenario a ``far away" celestial sphere is the radiation source. The integration stops when the photon reaches either the BH or the celestial sphere. In this case the disk is taken to be transparent. The numerical infinity (or emitting celestial sphere) setup is similar to the one described in \cite{Cunha:2018uzc}, being endowed with a two colour pattern to more easily identify the final position of the light rays on that sphere. Concretely, the celestial sphere is divided into two hemispheres, respectively above and below the equatorial plane, with each hemisphere being endowed with a colour: red (blue) for the North (South) hemisphere. The point in the celestial sphere immediately behind the BH, from the observer perspective, is attributed a white colour. For this scenario, the visualisation of a Schwarzschild BH surrounded by a test accretion disk, with inner edge at the ISCO is shown in Fig.~\ref{fig:scenarios} (right panel).

\end{description}

In both scenarios, the celestial sphere surrounds the BH, (part of) the disk and the observer. At the observer's position, we setup a local observer basis $\{\hat{e}_{(t)},\hat{e}_{(\rho)} ,\hat{e}_{(z)} ,\hat{e}_{(\varphi)}\}$ analogous to \cite{Cunha_1605}, where the photons' initial momentum is parameterised  by two angles~\cite{Cunha_1605}, $\alpha$ (vertical) and $\beta$ (horizontal). The results presented in the next section were obtained by discretising the angles $\alpha$ and $\beta$ in two arrays of 1000 values each, with both angles set in the interval $[-\tan^{-1}(10/15) \, ; \tan^{-1}(10/15)]\simeq [-0.6,0.6]$. The celestial sphere has a perimetral radius \cite{Cunha_1605} of $\tilde{r} = 30 M_{\rm total}$. Some remarks regarding the impact of this choice on the results are made at the end of the next section. Moreover, the observer is placed at the disk plane in both scenarios, with a coordinate $\rho_{\rm obs}$ obtained by solving the (perimetral radius) equation $\tilde{r}_{\rm obs}=\sqrt{g_{\varphi\varphi}(\rho_{\rm obs},z_{\rm obs})} = 15 M_{\rm total}$, with $z_{\rm obs}=0$.

The images in Fig.~\ref{fig:scenarios} correspond to the lensing and shadow of a Schwarzschild BH, under the two aforementioned scenarios.  When considering a heavy backreacting accretion disk,  in the next section, these images will be deformed. Thus, let us summarise the main features of the  images in Fig.~\ref{fig:scenarios}.

Consider first the left panel of Fig.~\ref{fig:scenarios}, starting from the outermost features. The large white annular region corresponds to light rays that are bent by the BH and hit (part of) the backward half of the accretion disk, with respect to the observer. It is a perfect double copy, north-south symmetric, since the observation is done from the equatorial plane. The forward half of the accretion disk (on the observer's side) is not seen covering the shadow, unlike the case of the non-equatorial observation in Fig.~\ref{fig0}. Indeed, for an equatorial observation, only a measure zero set of photons -- those with precisely vanishing $z$-momentum -- hit the forward section of the disk. The smaller grey annular region corresponds to light rays that are bent more strongly by the BH so that they cross the equatorial plane inside the inner edge of the disk, ending up at (the numerical) infinity. This produces an inverted copy of part of the sky: the north (south) hemisphere would be seen in the bottom (top) part of grey annulus. Light rays that approach even more the Schwarzschild photonsphere are bent even more and can end up in the forward half of the accretion disk. This corresponds to the tiny white ring around the central black region. An infinity succession of further, increasingly thinner, grey and white annular regions would be revealed by an infinite resolution, corresponding to light rays that skim, increasingly closer, the photonsphere. The black region is the BH shadow, corresponding to photons with an impact parameter smaller than that of the photonsphere ($\sqrt{27}M$).

The analysis of the right panel of Fig.~\ref{fig:scenarios} is similar. In particular, the aforementioned north-south inversion in the grey annular region of the left panel becomes now manifest. In fact, such inversion occurs in a larger angular region, which in the first scenario is partly covered by the disk. The main qualitatively new feature is the lensing of the white dot precisely behind the BH, with respect to the observer, into an Einstein ring.  One can see its angular size coincides with the outer boundary of the large white annulus in the left panel. Thus, this outer boundary corresponds to a single point of the outer edge of the disk (which is  at infinity): the point along the line of sight of the observer and the BH.  For more on the lensing by a Schwarzschild BH see, e.g.~\cite{Bohn2014}.

\section{Results}
\label{section:Results}

\subsection{First scenario - emission from the disk}\label{section:Shadow_case}

\begin{figure}[!h]
  \begin{subfigure}{.33\textwidth}
    \centering
    \includegraphics[width=0.9\linewidth]{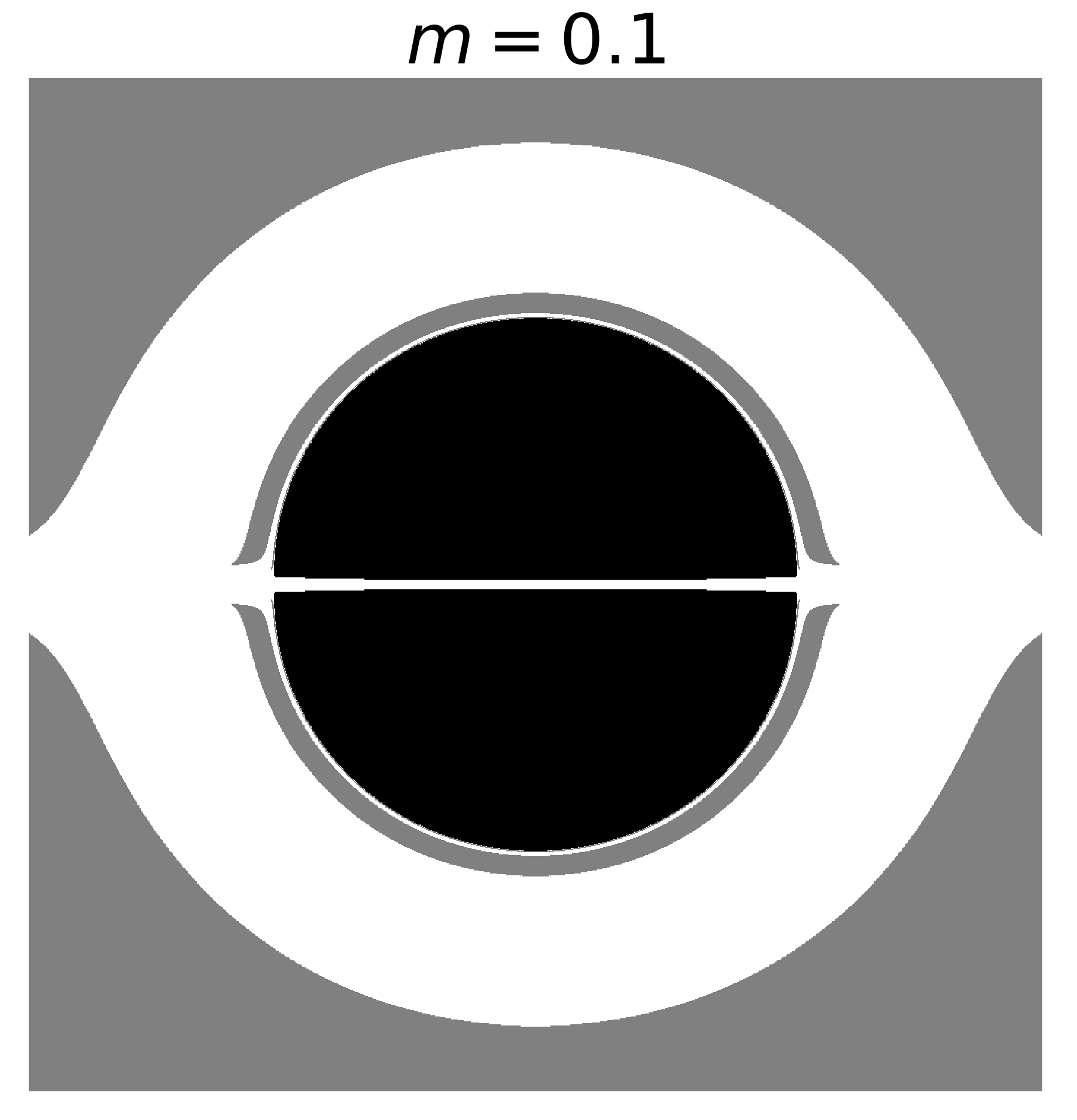}
    \caption{  }
  \end{subfigure}
   \begin{subfigure}{.33\textwidth}
    \centering
    \includegraphics[width=0.9\linewidth]{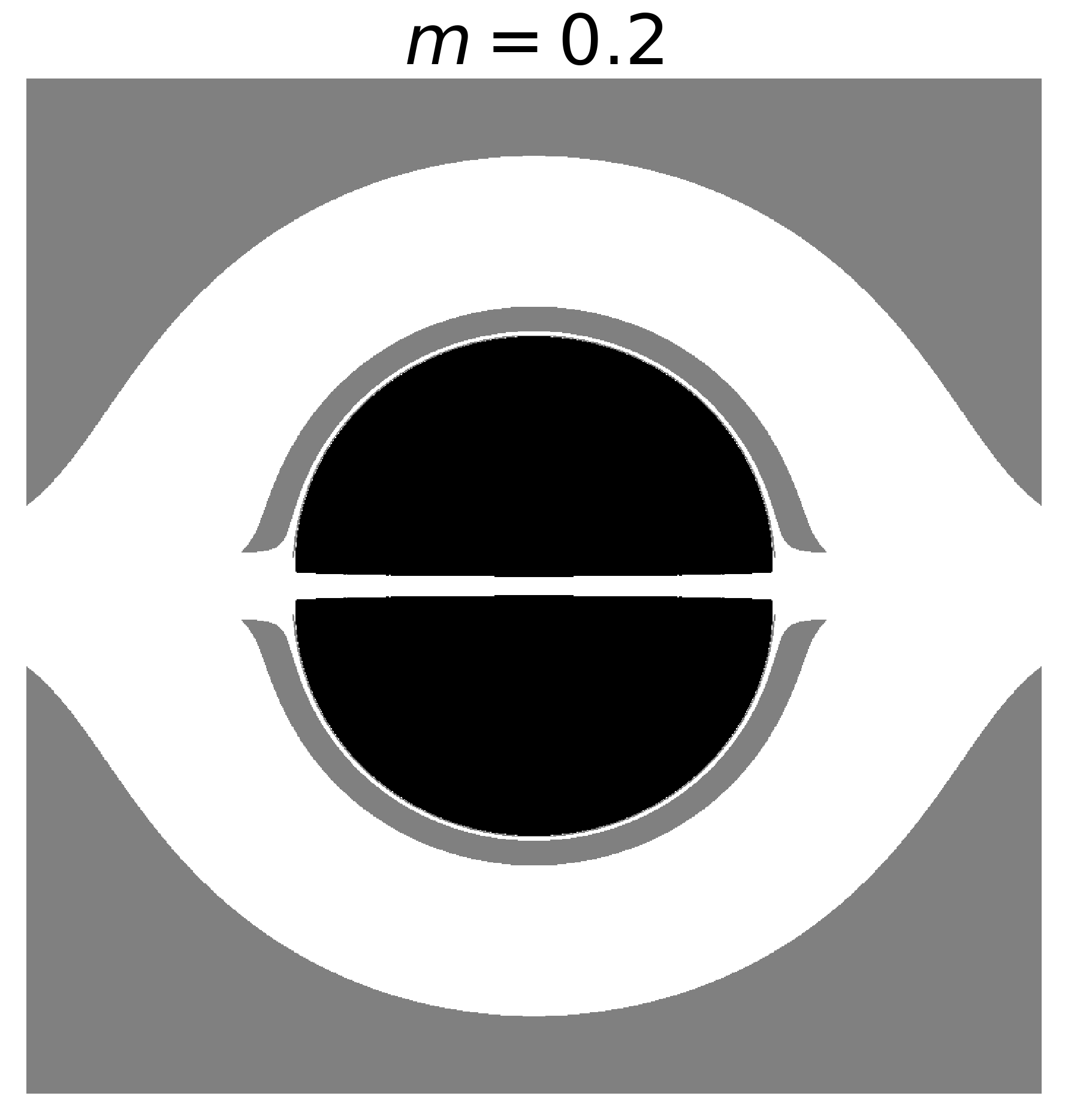}
    \caption{  }
  \end{subfigure}
   \begin{subfigure}{.33\textwidth}
    \centering
    \includegraphics[width=0.9\linewidth]{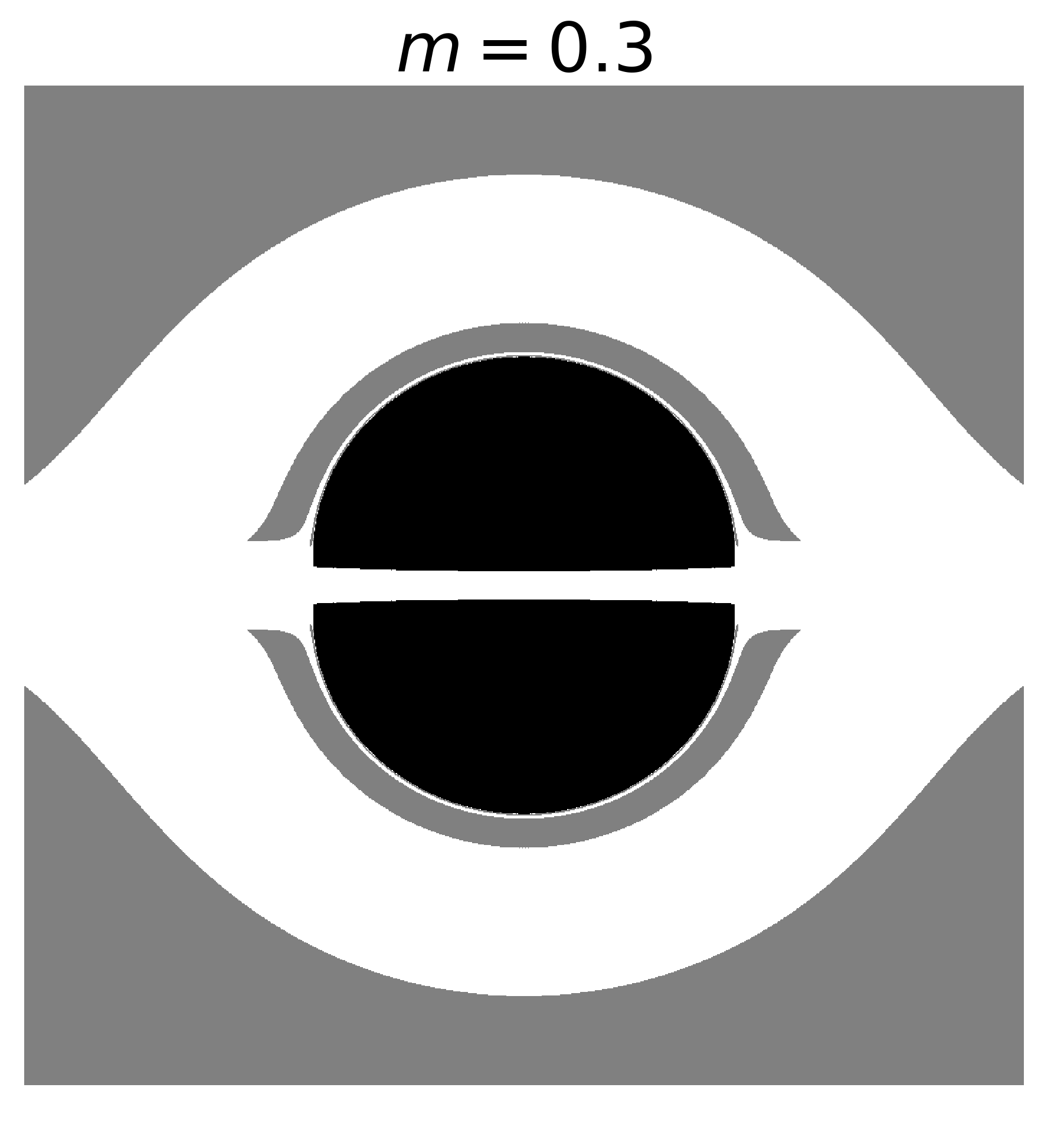}
    \caption{  }
  \end{subfigure}
    \begin{subfigure}{.33\textwidth}
    \centering
    \includegraphics[width=0.9\linewidth]{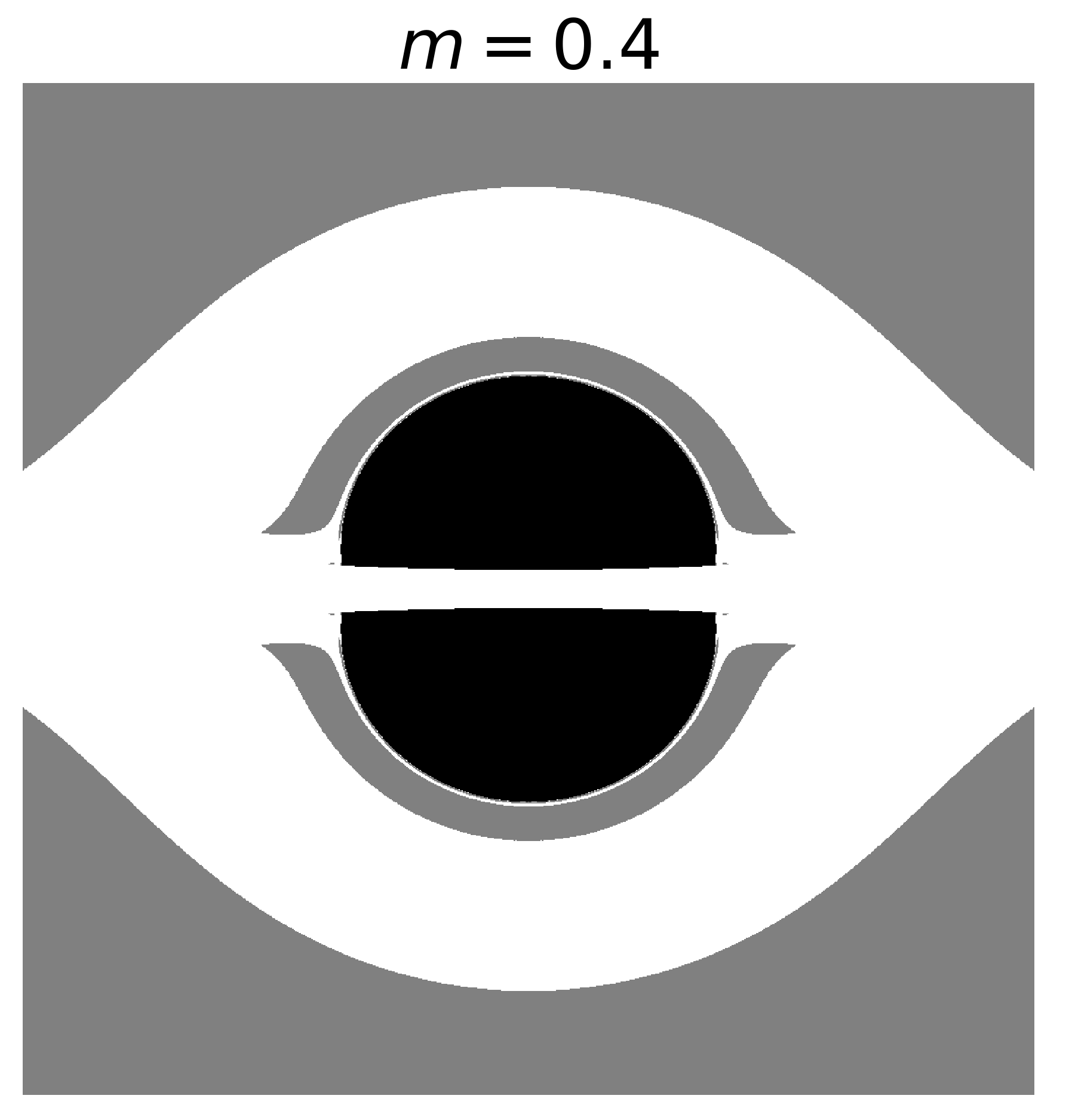}
    \caption{  }
  \end{subfigure}
  \begin{subfigure}{.33\textwidth}
    \centering
    \includegraphics[width=0.9\linewidth]{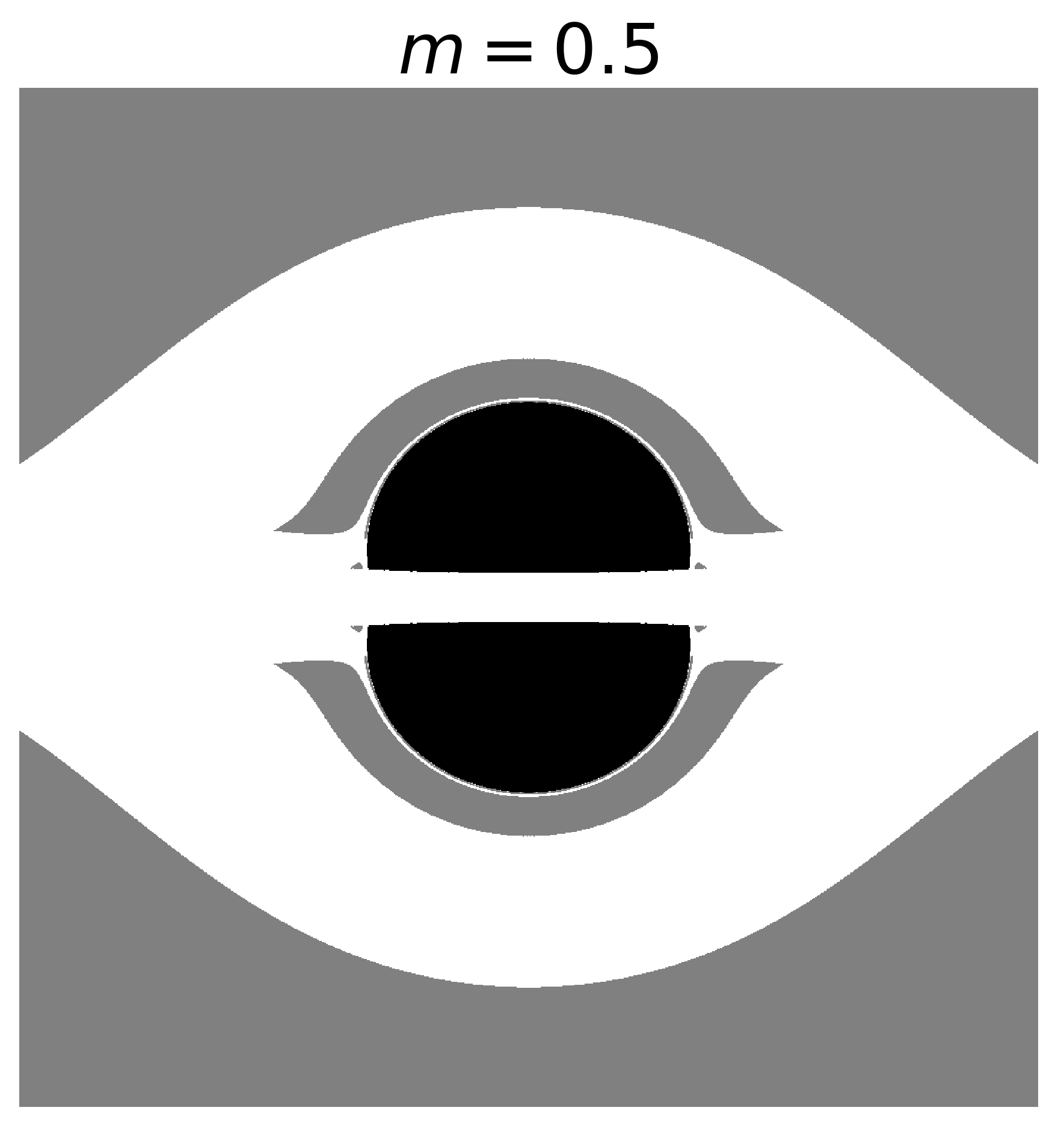}
    \caption{  }
  \end{subfigure}
   \begin{subfigure}{.33\textwidth}
    \centering
    \includegraphics[width=0.9\linewidth]{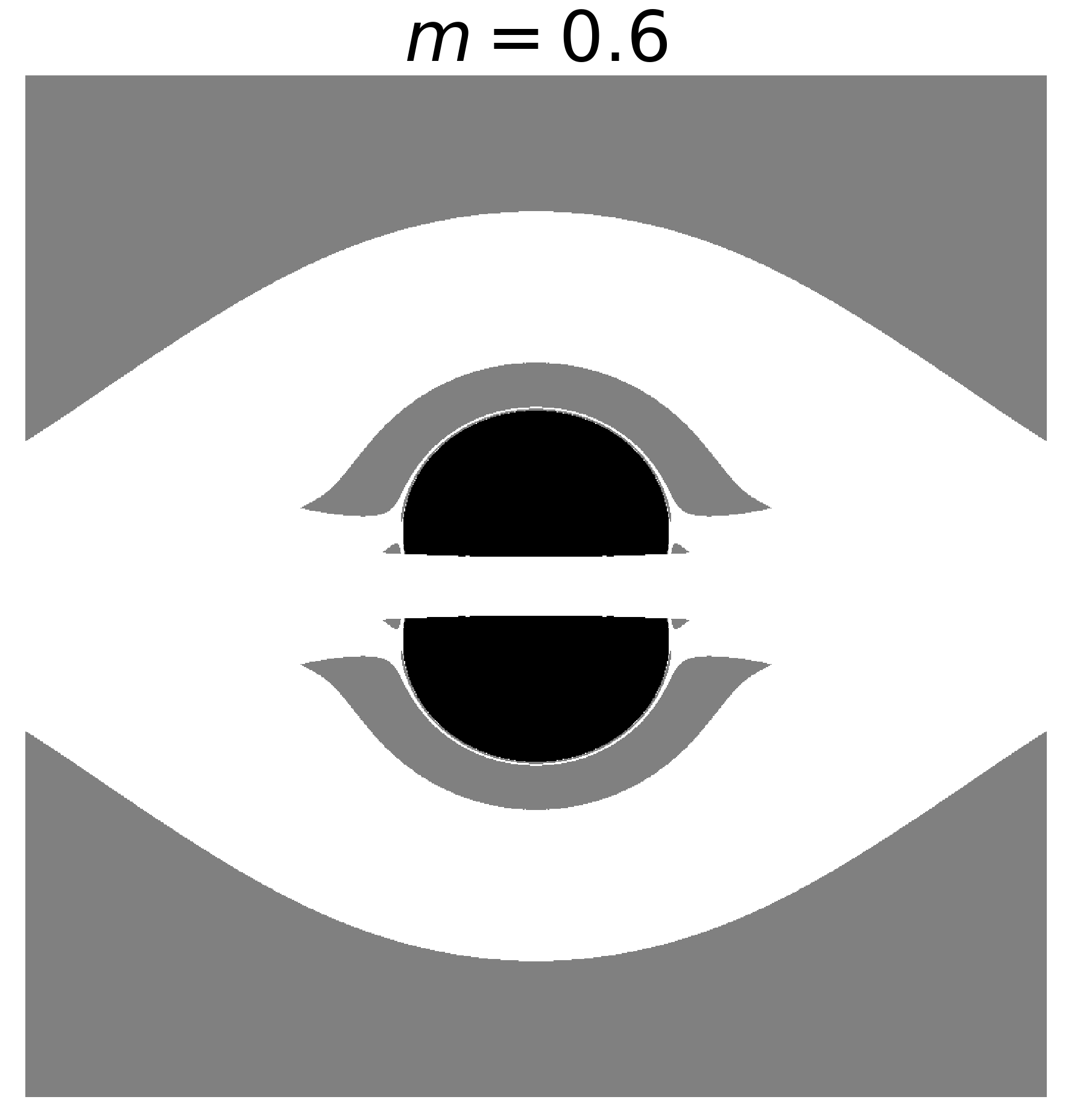}
    \caption{  }
  \end{subfigure}
  \caption{\small Imaging of the BH+LL disk system with fixed $a = 3.0\;$ and varying $m$ when illuminated from the disk: a) $m=0.1$, b) $m=0.2$, c) $m=0.3$, d) $m=0.4$, e) $m=0.5$, f) $m=0.6$. The $\alpha$ (vertical axis) and $\beta$ (horizontal axis) ranges are $\alpha,\beta \in [-0.6,0.6]$.}
  \label{fig:Shadow_disk_a}
\end{figure}

In Fig.~\ref{fig:Shadow_disk_a} we exhibit the visualisation of the BH+LL disk system illuminated from the disk. We fix $a=3.0$ and vary $m$, in accordance to the sequence of points in Fig.~\ref{fig:AllowedRegions}. The sequence of images
starts with the smallest $m=0.1$ (left top panel).  In this case the forward part of the disk (in between the observer and the BH horizon) becomes visible, unlike Fig.~\ref{fig:scenarios} (left panel), covering a thin slice of the shadow. This means there is now an open set of initial conditions (photons with small, but non-vanishing $z$-momentum) that are bent towards the disk in between the observer and the BH. This is a main novel feature of a heavy backreacting disk: its gravitational pull attracts photons.

\begin{figure}[h!]
\begin{center}
\includegraphics[width=0.71\textwidth]{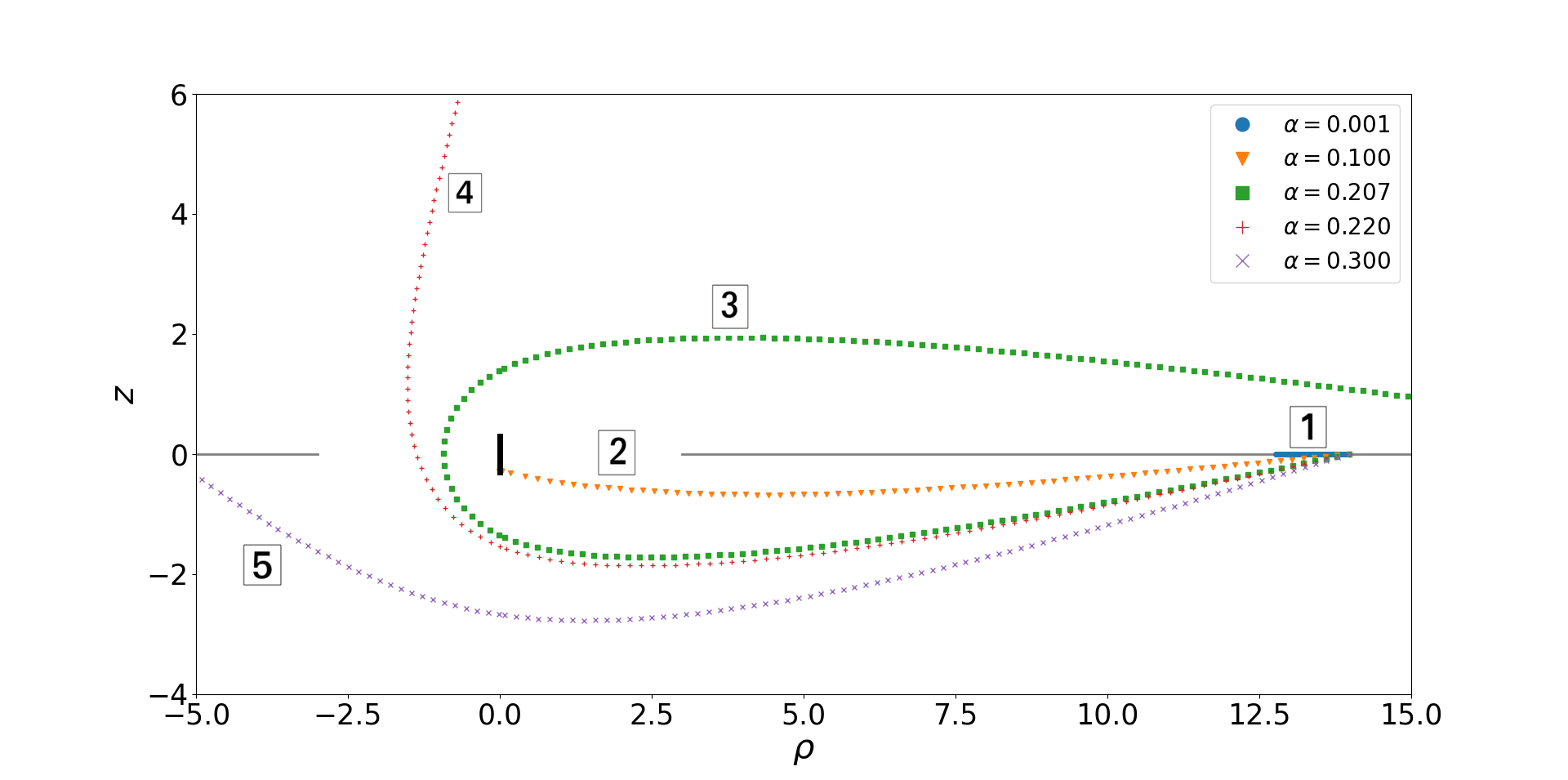}\hspace{-0.9cm}{\includegraphics[width=0.34\textwidth]{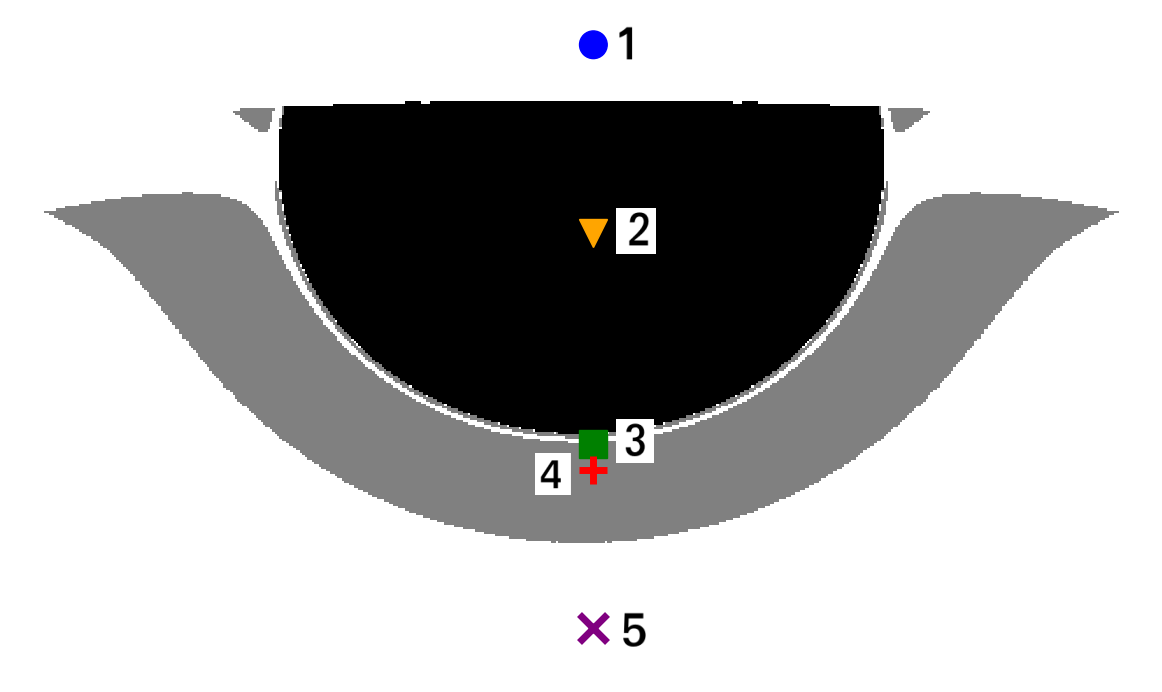}}
\caption{\small (Left panel) $\rho-z$ trajectory of some illustrative photons in panel (f) of Fig.~\ref{fig:Shadow_disk_a} - $m=0.6$ and $a=3.0$. The chosen photons have $\beta \approx 0.006$ and different values of $\alpha$ (labels 1 to 5). The negative values of $\rho$ correspond to the back side of the BH, with respect to the observer, which is located at $z_{\rm obs}=0$ and $\rho_{\rm obs}\simeq 14$. The BH location is represented by the black vertical line at $\rho=0$, whereas the disk is depicted by the grey horizontal line at $z=0$ and $|\rho|\geqslant 3$. (Right panel) The trajectories exhibited in the left panel are associated to points in a zoom of panel (f) of Fig.~\ref{fig:Shadow_disk_a}. The image shows two small ``earlobes", on each side of the shadow, near the equatorial plane. These peculiar structures are discussed in the Appendix.}
\label{trajectory1}
\end{center}
\end{figure}

The section of the disk behind the BH is seen, again, above and below the BH shadow, due to light bending. But the corresponding white annulus now opens up at the edges, due to light rays with large impact parameter that, nonetheless, can still fall towards large $\rho$ sections of the disk. The smaller grey annulus in Fig.~\ref{fig:scenarios} (left panel) also opens up now near the equator, for a similar reason: light rays with small $z$-momentum fall onto the disk.  In between the shadow and the thin grey annulus, a tiny white ring can still be observed, which, as in Fig.~\ref{fig:scenarios} (left panel), corresponding to photons that skim the fundamental photon orbits (i.e., the generalisation of the photonsphere for non-Schwarzschild BHs, see~\cite{Cunha:2017eoe}) and end up on the disk. In the right panel of Fig. \ref{trajectory1}, another layer of grey between the tiny white region and the shadow edge is still visible, corresponding to photons with impact parameters even closer to that of fundamental photon orbits and escape to (numerical) infinity. As in the unperturbed Schwarzschild case, one expects further white and grey regions to exist, which would be unveiled by increasing the resolution.

As $m$ increases we move from left to right, top to bottom in Fig.~\ref{fig:Shadow_disk_a}. One observes a qualitatively similar image to that just described but with two outstanding trends. Firstly, there is a progressive optical enlargement of the disk image covering part of the shadow (white band), despite the fact that the disk is infinitesimally thin. This is a consequence of the disk's increasing ``weight". Secondly, for larger $m$, the shape of the shadow varies from the familiar circular shape to a more prolate spheroid.

To illustrate the correspondence between points in the image and the behaviour of the corresponding light rays, some of the photons' trajectories are shown in Fig.~\ref{trajectory1}, for a fixed angle $\beta \approx 0$ (specifically, we have taken $\beta \approx  0.006$), but with different angles $\alpha$, labeled by the numbers $1$ to $5$. Depending on the corresponding angles $\alpha$, the photons' trajectories can end: up hitting the BH, thus belonging to the shadow (for $\alpha = 0.1$, label 2); at  (numerical) infinity (for $\alpha = 0.220$, label 4); or on the disk (the trajectories corresponding to the remaining angles, labels 1,3,5). The latter trajectories confirm that the white portions of the image correspond to different parts of the disk:
\begin{description}
\item[i)] The trajectory with $\alpha=0.001$ (label 1) hits the forward portion of the disk in between the observer and the BH. The white section in the middle of the shadow corresponds to the disk in between the observer and the BH;
\item[ii)] The trajectory with $\alpha=0.207$ (label 3) hits the portion of the disk behind the observer. This tiny white section in between the shadow edge and the grey annulus spans the forward portion of the disk, both behind the observer and between the observer and the BH;
\item[iii)] The trajectory with $\alpha=0.3$ (label 5) hits the portion of the disk behind the BH. In accordance with the description of Fig.~\ref{fig:scenarios} (left panel), the large white annular region corresponds to (part of the) backward disk.
\end{description}

\subsection{Second scenario - emission from the celestial sphere}\label{section:Lensing_case}
\begin{figure}[h!]
  \begin{subfigure}{.33\textwidth}
    \centering
    \includegraphics[width=0.9\linewidth]{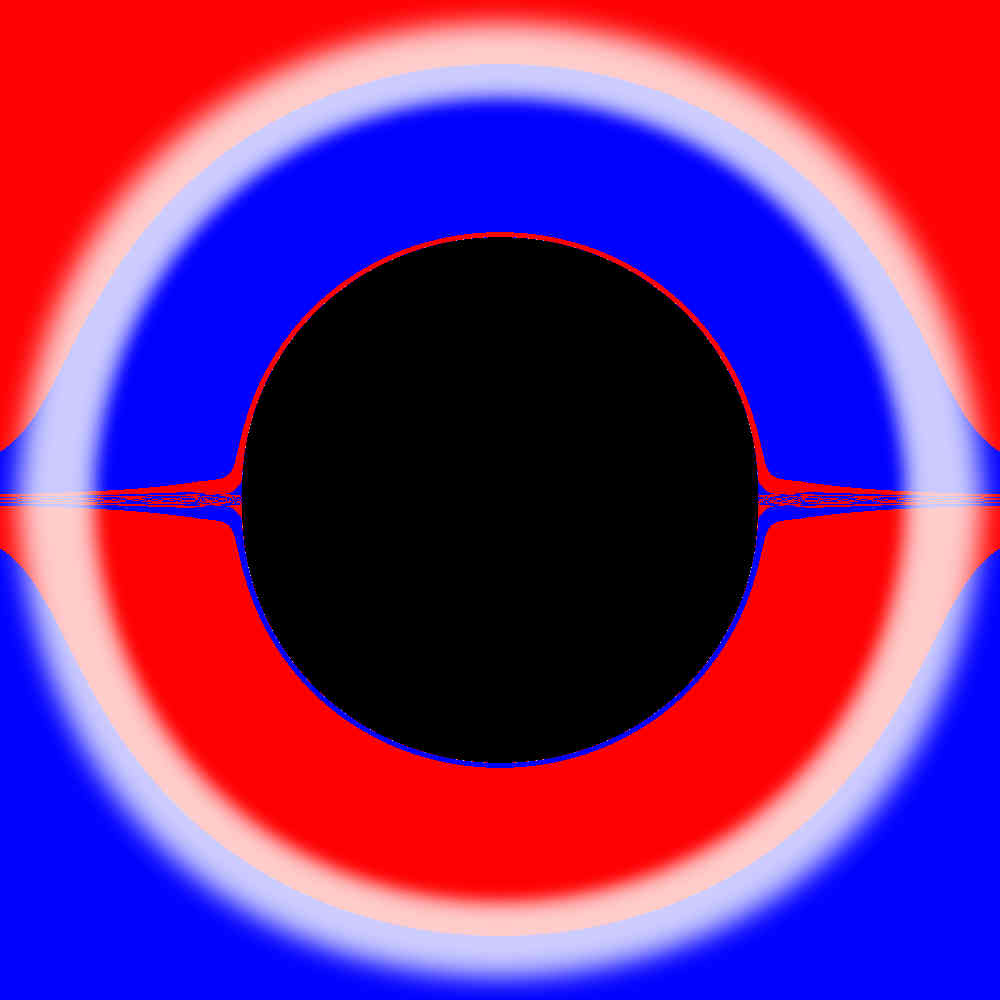}
    \caption{  }
  \end{subfigure}
   \begin{subfigure}{.33\textwidth}
    \centering
    \includegraphics[width=0.9\linewidth]{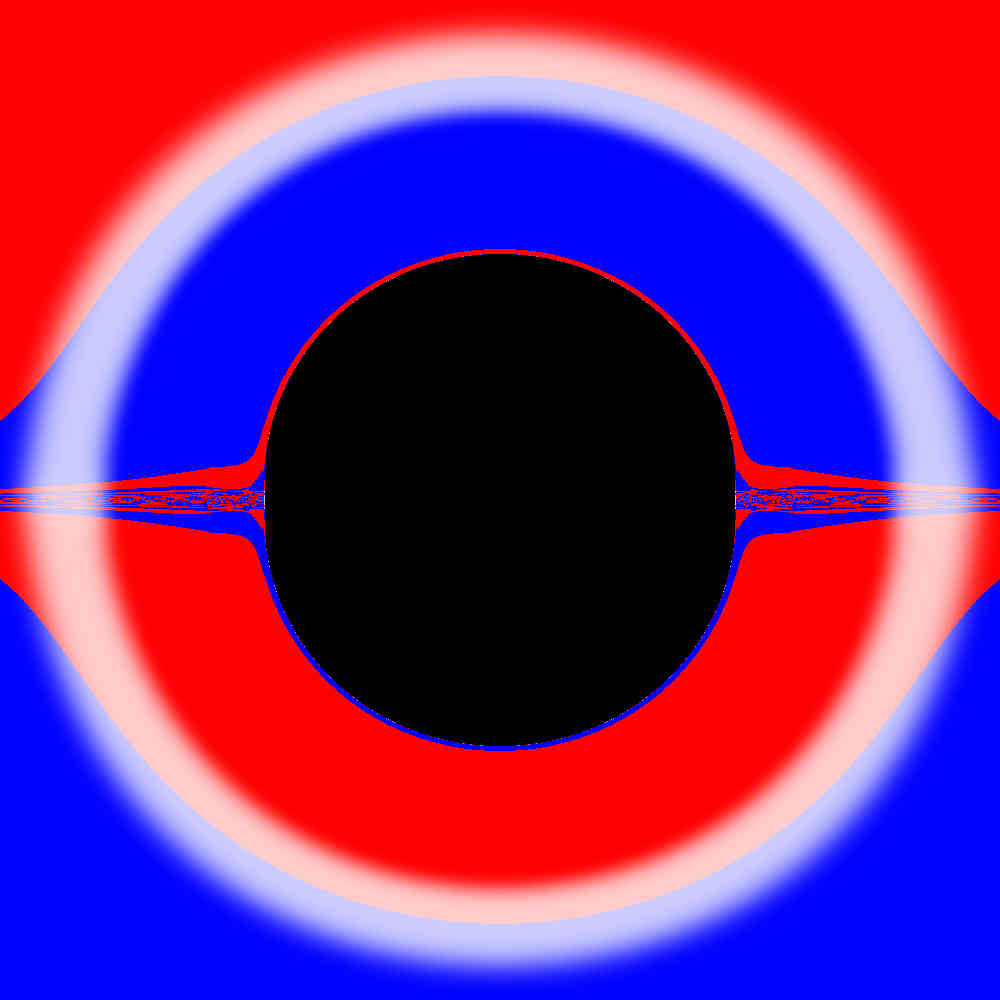}
    \caption{  }
  \end{subfigure}
   \begin{subfigure}{.33\textwidth}
    \centering
    \includegraphics[width=0.9\linewidth]{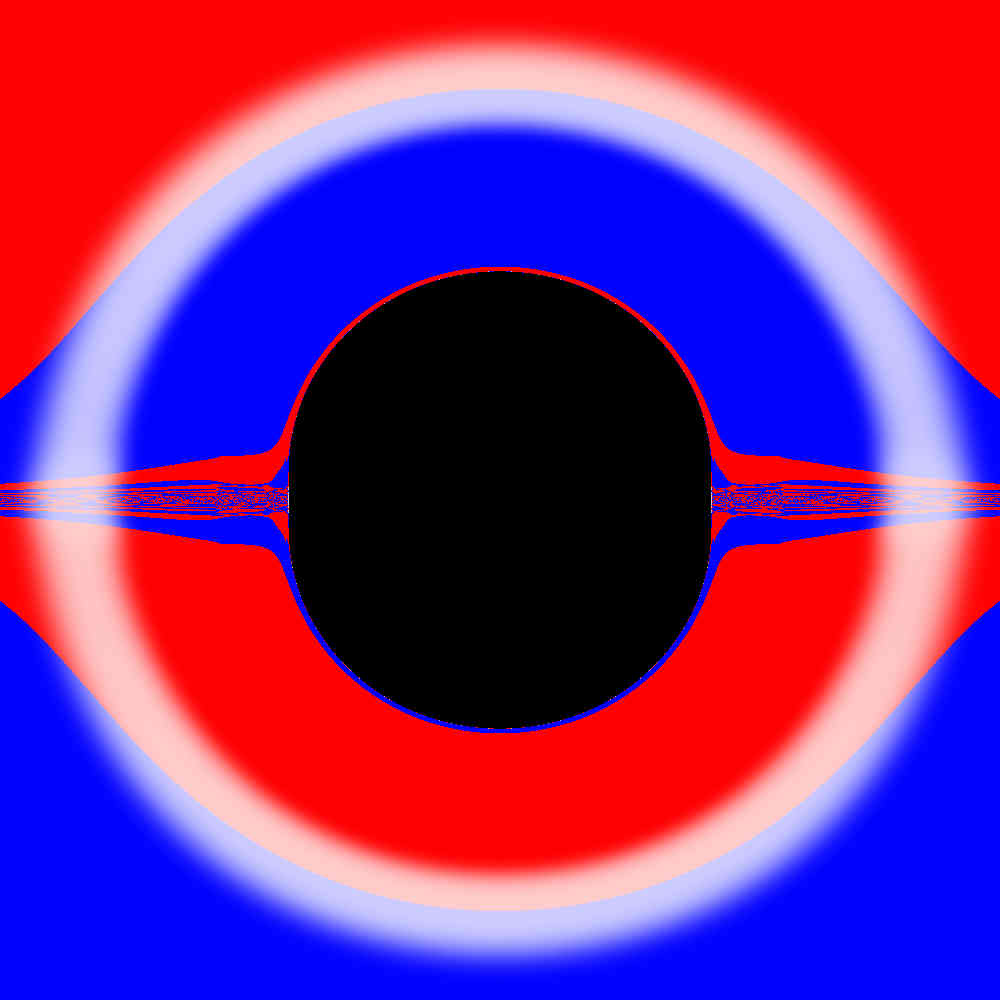}
    \caption{  }
  \end{subfigure}
   \begin{subfigure}{.33\textwidth}
    \centering
    \includegraphics[width=0.9\linewidth]{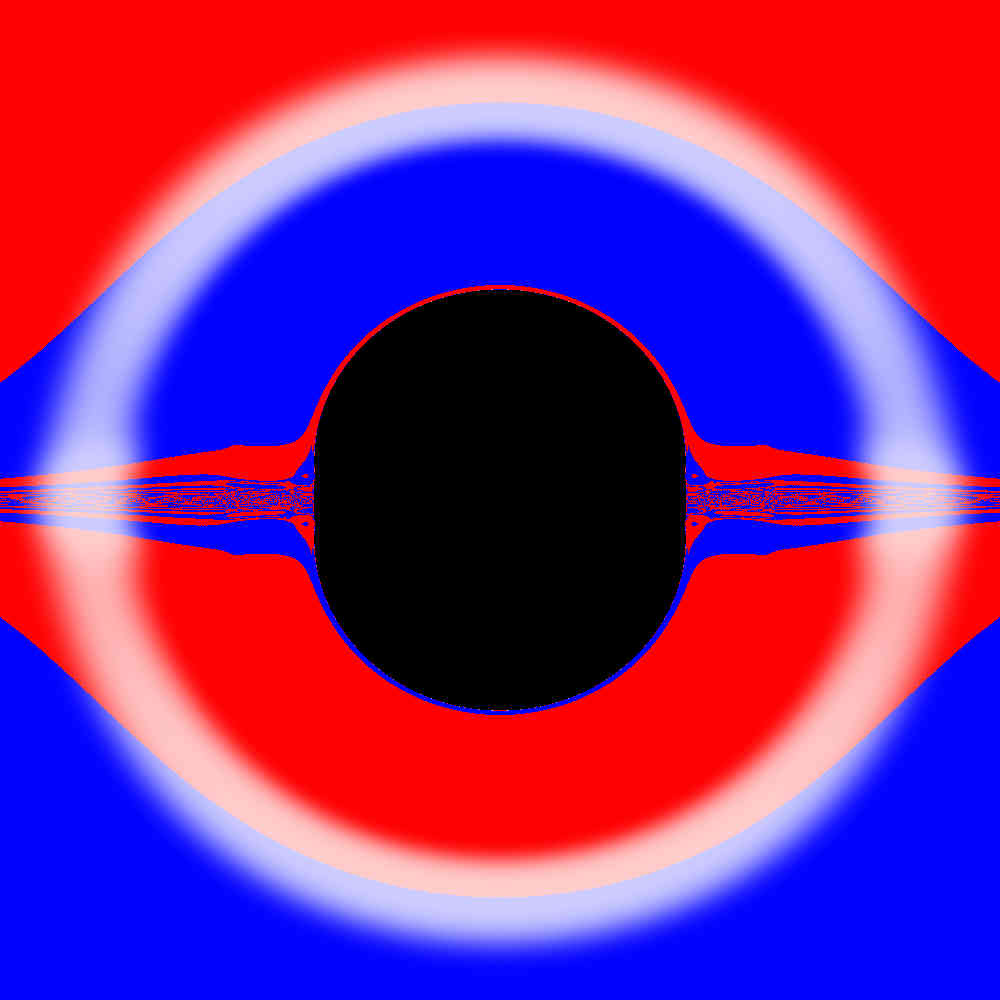}
    \caption{  }
  \end{subfigure}
   \begin{subfigure}{.33\textwidth}
    \centering
    \includegraphics[width=0.9\linewidth]{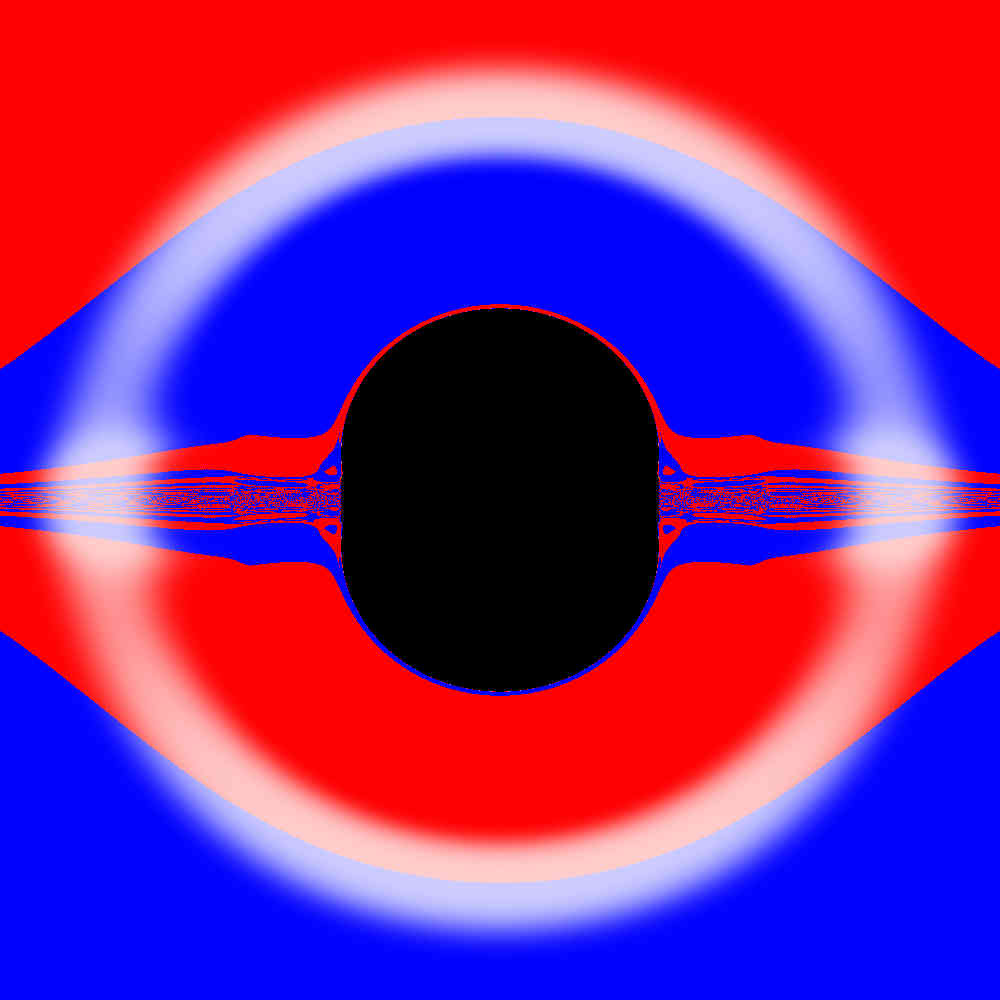}
    \caption{  }
  \end{subfigure}
    \begin{subfigure}{.33\textwidth}
    \centering
    \includegraphics[width=0.9\linewidth]{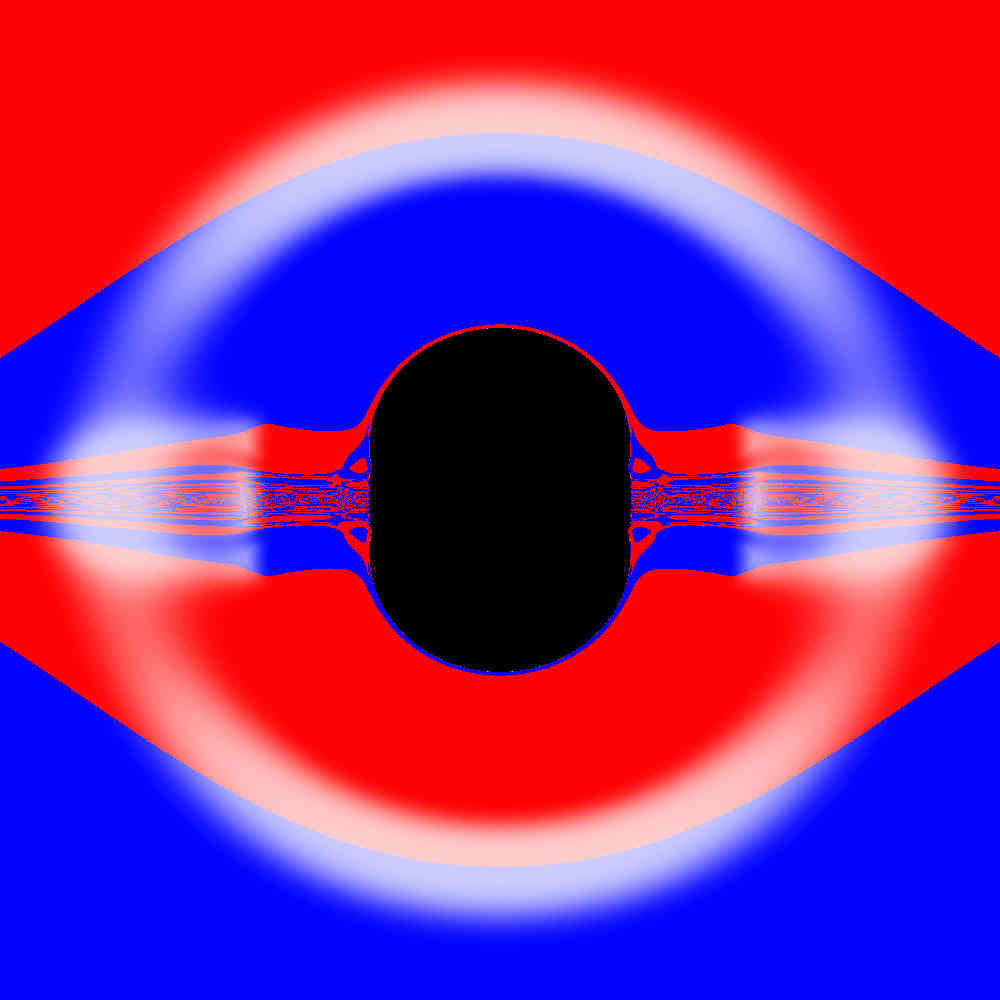}
    \caption{  }
  \end{subfigure}
    \caption{\small Imaging of the BH+LL disk system with fixed $a = 3.0\;$ and varying $m$ when illuminated from the far away celestial sphere: a) $m=0.1$, b) $m=0.2$, c) $m=0.3$, d) $m=0.4$, e) $m=0.5$, f) $m=0.6$. Again, the $\alpha$ (vertical axis) and $\beta$ (horizontal axis) ranges are $\alpha,\beta \in [-0.6,0.6]$. Red (blue) colour represent an endpoint in the North (South) hemisphere of the celestial sphere.}
  \label{fig:Lensing_disk_a}
\end{figure}

In Fig.~\ref{fig:Lensing_disk_a} we exhibit the visualisation of the BH+LL disk system illuminated from a far away celestial sphere. Again, we fix $a=3.0$ and vary $m$, following the sequence of points in Fig.~\ref{fig:AllowedRegions}. In this case we consider the disk transparent: the photons do not stop on the disk, but rather pass through it.

A few words about the transmission of the photons across the disk, due the discontinuity of the $z$-derivatives of the metric. Clearly, the velocity is continuous and the acceleration is discontinuous. The problem is analogous to a classical mechanics setup where a ball rolling through a horizontal plane finds a slope, with the crossover being cuspy. The acceleration has a discontinuity at the cusp; but the velocity is continuous. Our case is similar: the 4-velocity of the photons through the disk is continuous. Thus, one applies the same 4-velocity immediately above and below the disk, but with a different acceleration, given by the different connection components above and below.

\begin{figure}[h!]
\begin{center}
\includegraphics[width=0.65\textwidth]{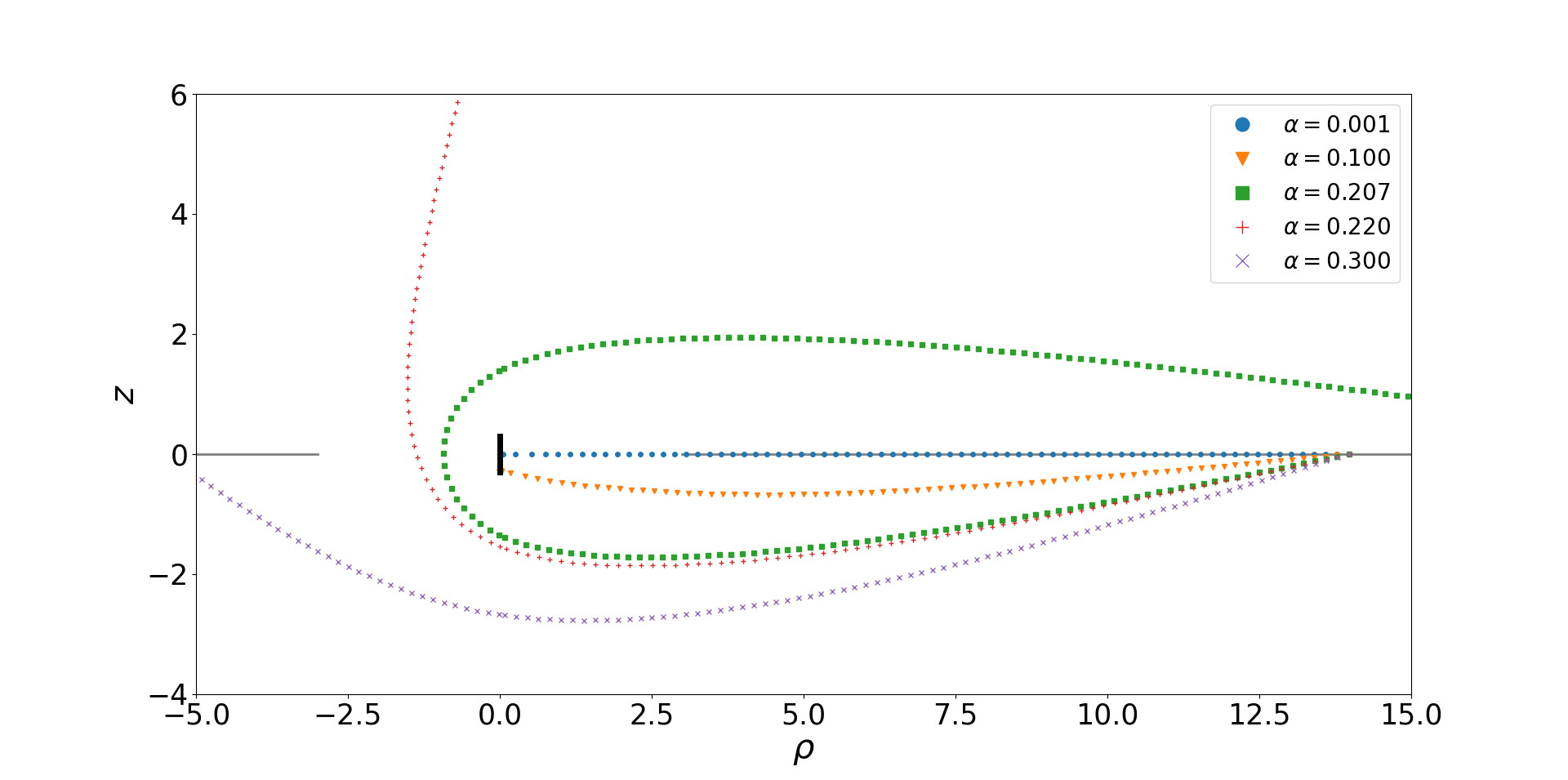}\vspace*{-0.1cm}
\includegraphics[width=0.65\textwidth]{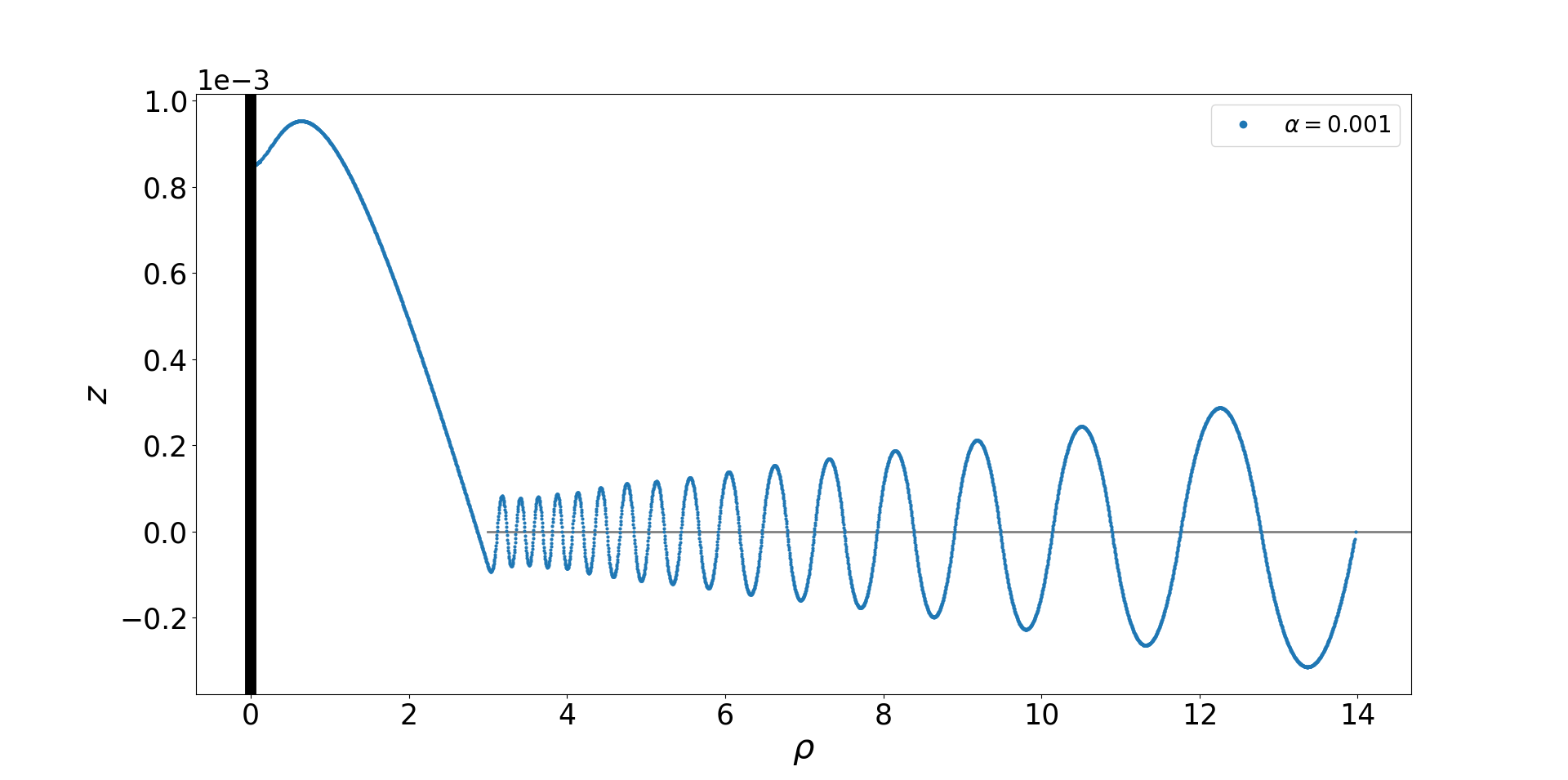}
\caption{\small (Top panel) A similar $\rho-z$  representation as in Fig.~\ref{trajectory1} but now for some illustrative photons in panel (f) of Fig.~\ref{fig:Lensing_disk_a} - $m=0.6$ and $a=3.0$. The chosen photons have $\beta \approx 0$ and different values of $\alpha$. Unlike Fig.~\ref{trajectory1}, photons can now cross the disk. Thus, the trajectory with $\alpha=0.001$ does not stop at the first disk crossing; it continues until it falls into the horizon. (Bottom panel) Zoom of the trajectory with $\alpha=0.001$. The light ray gets trapped by the disk's gravitational potential, oscillating around it; the amplitude of the oscillations decreases, together with the distance between nodes, as $\rho$ decreases, since the disk density's is increasing for smaller $\rho$. At $\rho=3$, i.e., at the inner edge of the disk (horizontal grey line),  the photon trajectory stops oscillating and falls into the BH (represented by the vertical black line).}
\label{trajectory2}
\end{center}
\end{figure}

In Fig.~\ref{fig:Lensing_disk_a}, the sequence of images starts again with $m=0.1$ (top left panel).  The new features, as compared to Fig.~\ref{fig:scenarios} (right panel), occur near the equatorial plane. Firstly, the BH shadow is not covered by the disk any longer. The new behaviour of the photons that were falling in the forward section of the disk, under the first scenario, is illustrated in Fig.~\ref{trajectory2}.

\begin{figure}[h!]
\begin{center}
\includegraphics[width=0.8\textwidth]{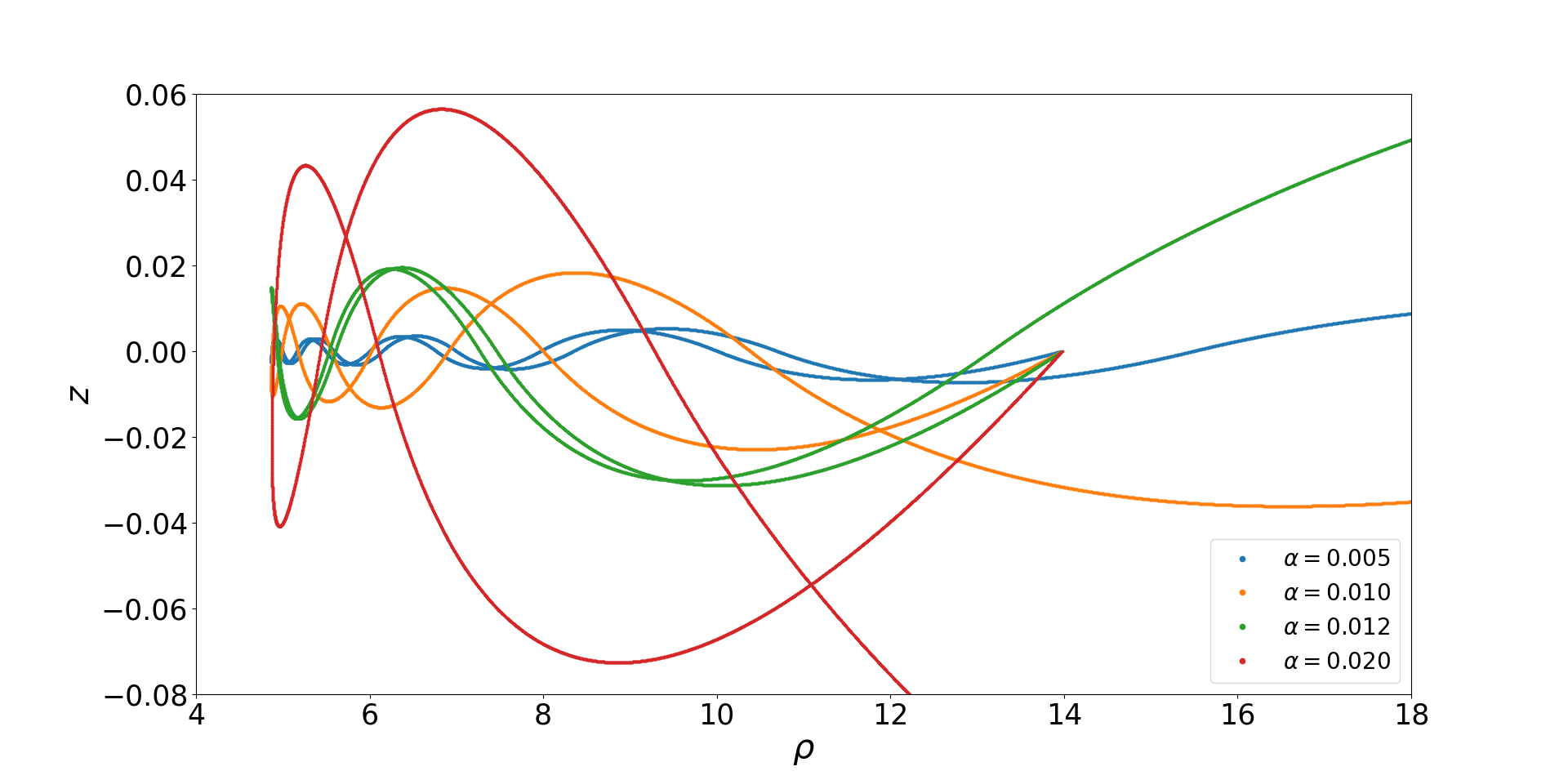}
\caption{\small $\rho-z$ trajectory of some illustrative photons in panel (f) of Fig.~\ref{fig:Lensing_disk_a} - $m=0.6$ and $a=3$. The chosen photons have $\beta= 0.45$ and different values of $\alpha$. All trajectories have an initial negative $z$ momentum, when departing from the observation point (at $\rho\simeq 14$) in backwards ray-tracing. Thus, they correspond to looking ``down", i.e., towards the southern hemisphere. Nonetheless, depending on the number of oscillations, two trajectories end up in the north hemisphere (blue, $\alpha=0.005$ and green $\alpha=0.12$) whereas two others in the south hemisphere (orange, $\alpha=0.010$ and red, $\alpha=0.020$), as they move towards large $\rho$, after reaching a minimum value of $\rho$. So, progressively looking down, the observer sees, alternately, north, south, north, south. }
\label{trajectory3}
\end{center}
\end{figure}

Fig.~\ref{trajectory2} shows how rays corresponding to a small angle $\alpha$ have a small $z$-component of their momentum and get swayed by the disk's potential. For small $\beta$ such light rays end up falling into the BH, thus becoming part of the shadow, as in Fig.~\ref{trajectory2} (bottom panel). For sufficiently large $\beta$ angle, however, the light rays still oscillate by virtue of the disk's gravity, but they do not fall into the BH; rather they escape to infinity, eventually. Depending on the number of these oscillations they may end up in the north or south hemisphere of the celestial sphere. This is illustrated in Fig.~\ref{trajectory3}.

\begin{figure}[h!]
  \begin{subfigure}{.33\textwidth}
    \centering
    \includegraphics[width=0.9\linewidth]{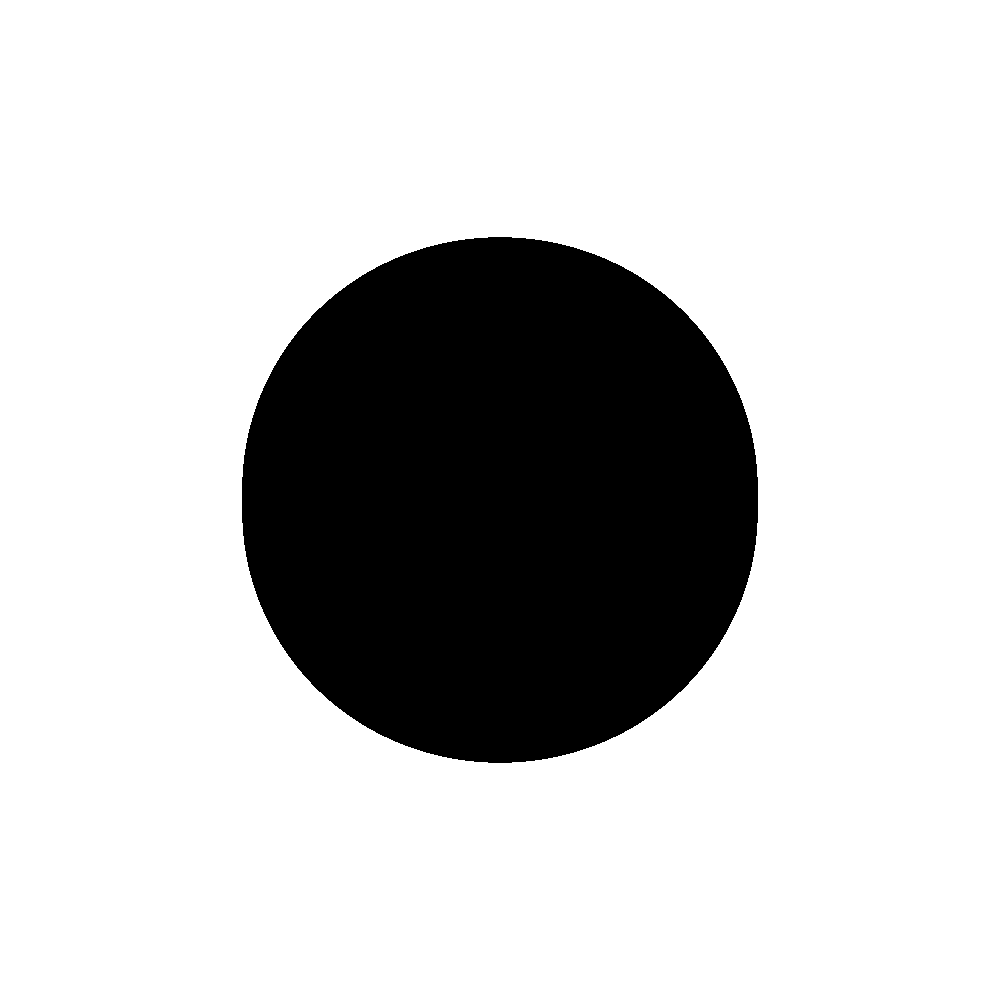}
    \caption{  }
  \end{subfigure}
   \begin{subfigure}{.33\textwidth}
    \centering
    \includegraphics[width=0.9\linewidth]{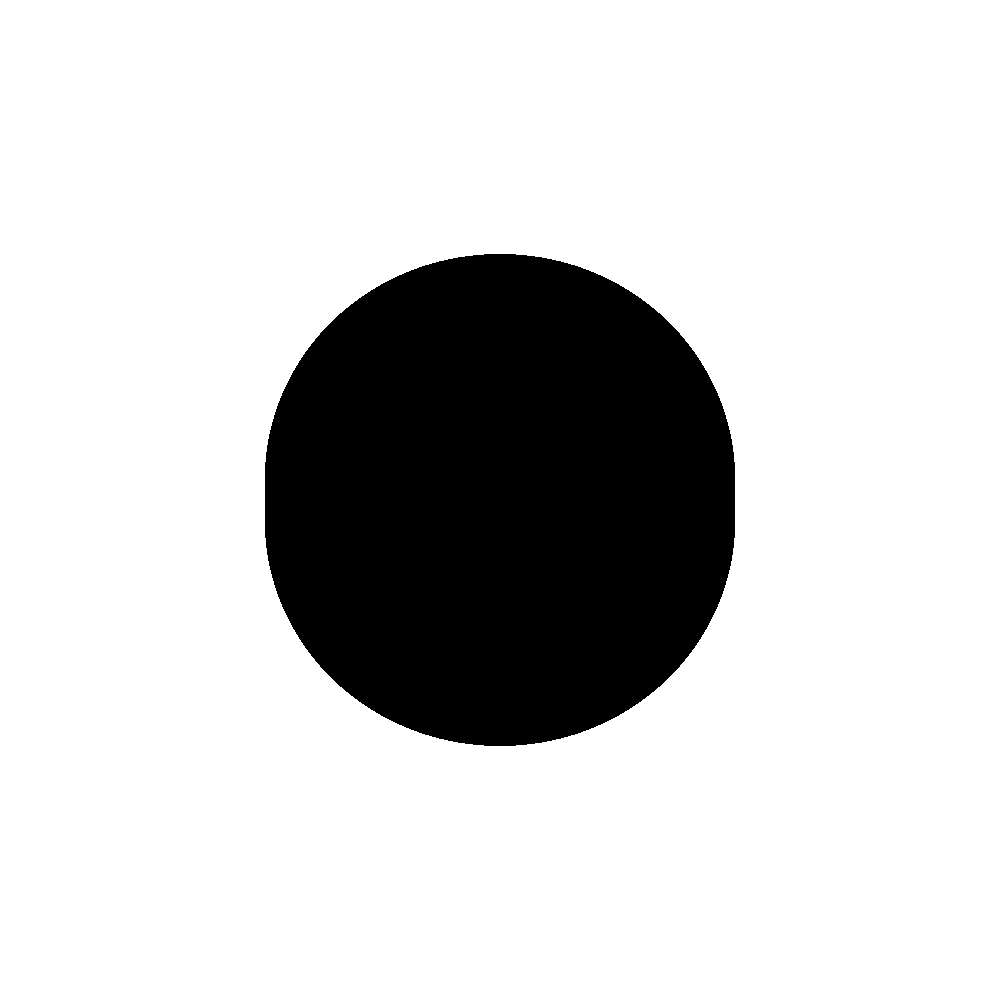}
    \caption{  }
  \end{subfigure}
   \begin{subfigure}{.33\textwidth}
    \centering
    \includegraphics[width=0.9\linewidth]{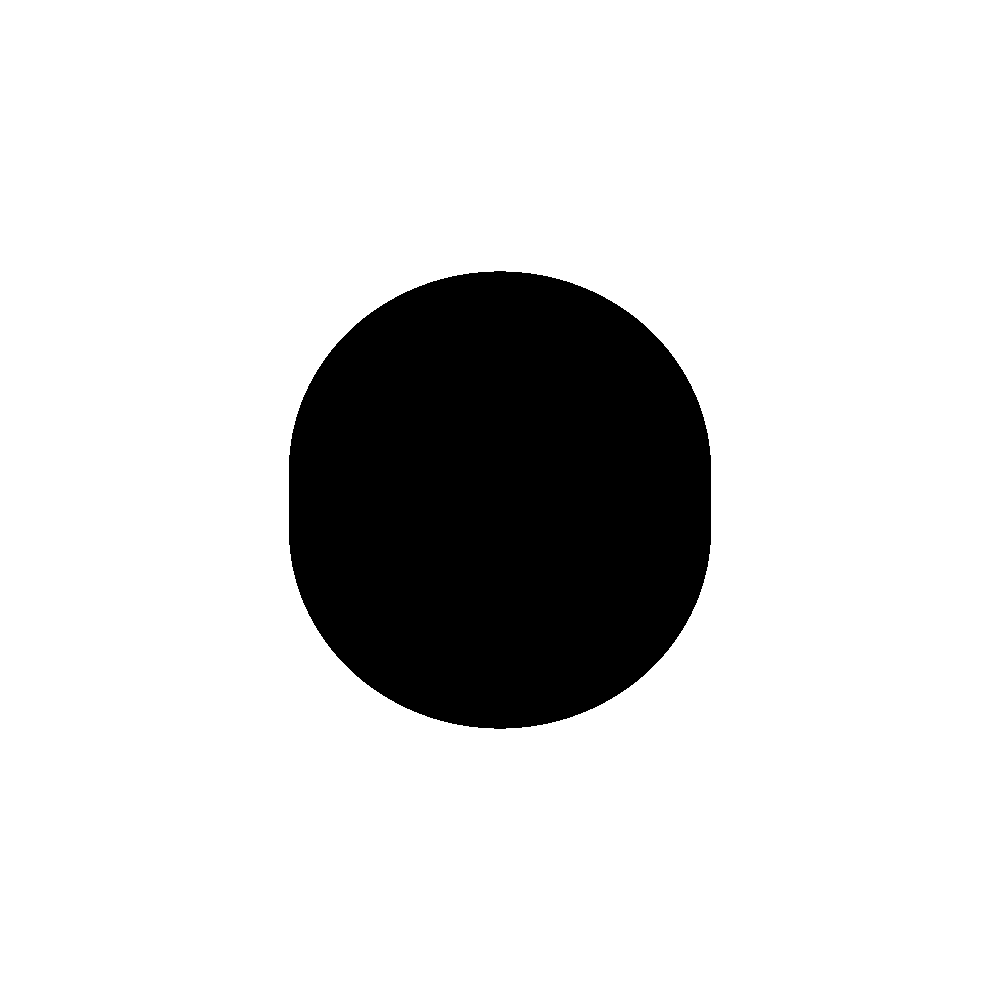}
    \caption{  }
  \end{subfigure}
   \begin{subfigure}{.33\textwidth}
    \centering
    \includegraphics[width=0.9\linewidth]{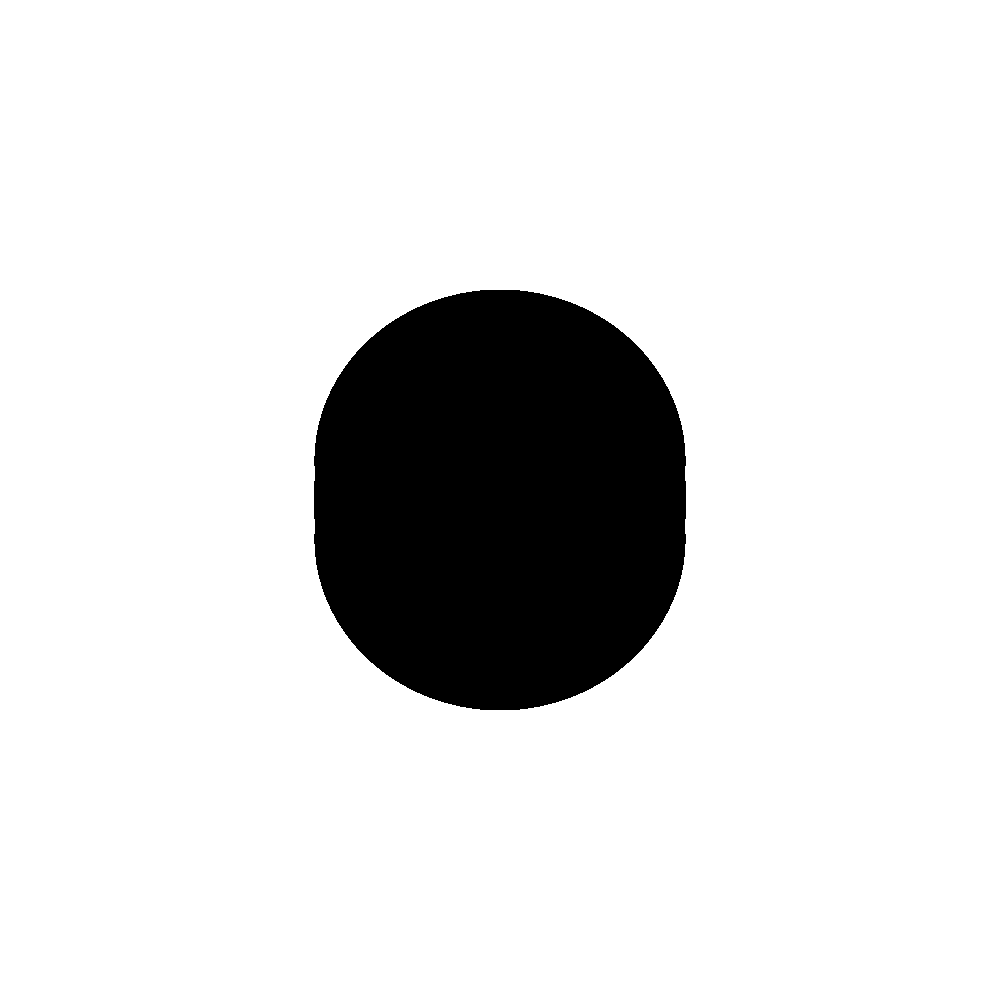}
    \caption{  }
  \end{subfigure}
   \begin{subfigure}{.33\textwidth}
    \centering
    \includegraphics[width=0.9\linewidth]{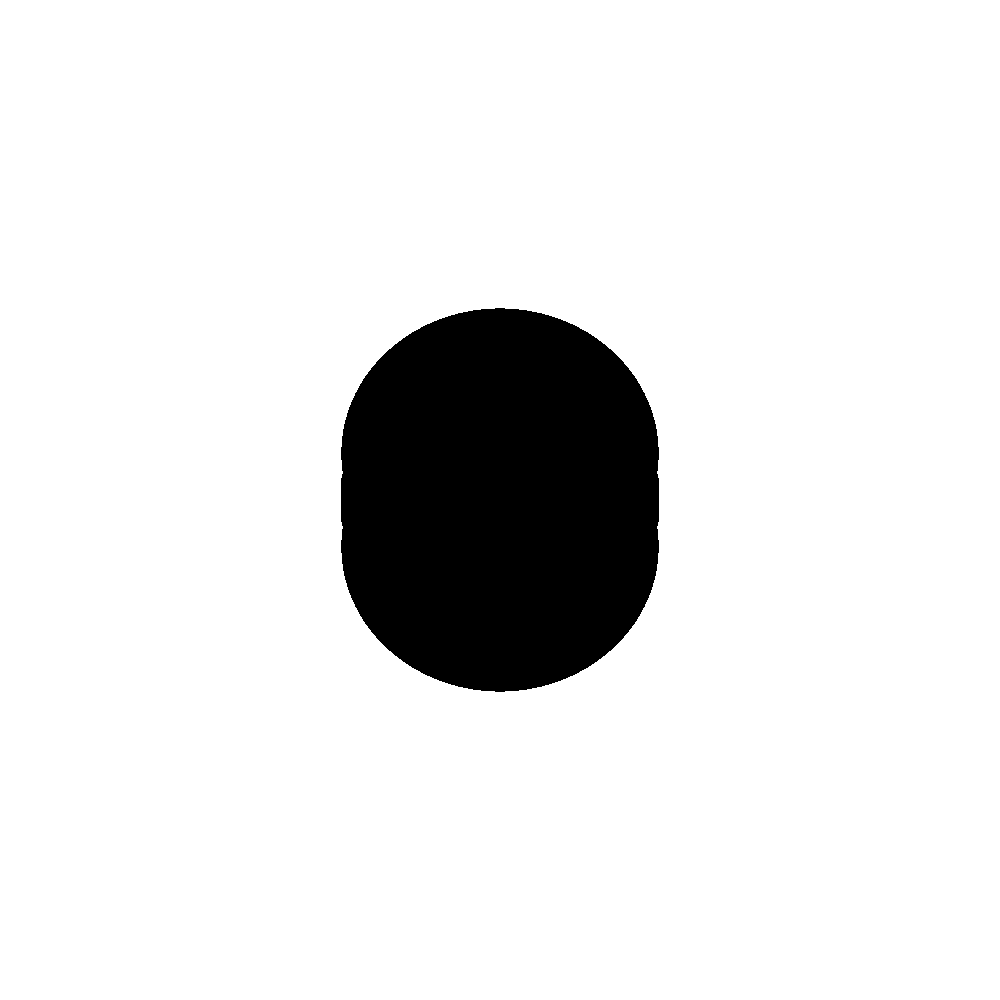}
    \caption{  }
  \end{subfigure}
    \begin{subfigure}{.33\textwidth}
    \centering
    \includegraphics[width=0.9\linewidth]{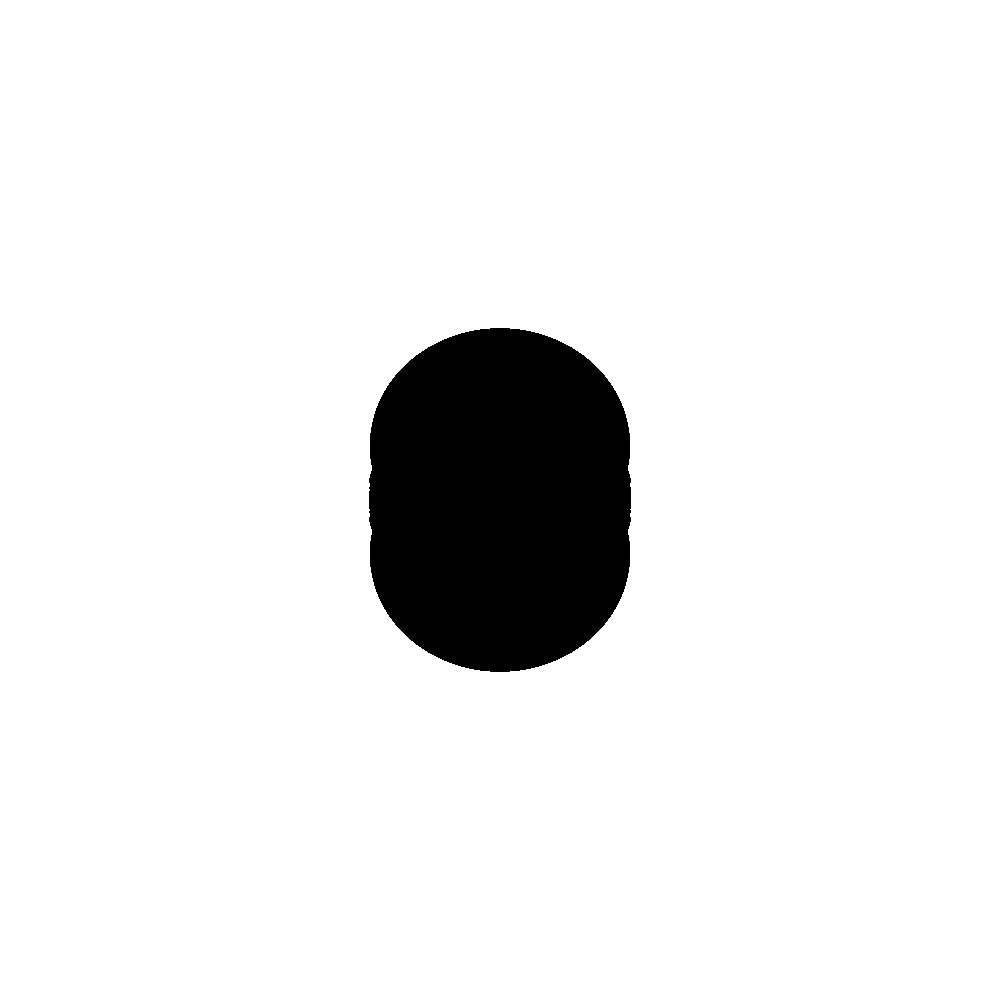}
    \caption{  }
  \end{subfigure}
    \caption{\small The BH shadow, without the lensing, in the images of Fig.~\ref{fig:Lensing_disk_a}.}
  \label{fig:shadow}
\end{figure}

The manifestation of this trapping by the disk potential outside the shadow but in the neighbourhood of the equatorial plane, i.e., for large enough $\beta$  and sufficiently small $\alpha$,  are the outwards, spiky features already seen in top left panel of Fig.~\ref{fig:Lensing_disk_a}. Close inspection reveals an irregular pattern of neighbouring pixels with different colours, i.e., chaotic behaviour. Increasing $m$, i.e., moving from left to right, top to bottom in Fig.~\ref{fig:Lensing_disk_a}, the spiky regions become thicker and the chaotic regions become clearer. These are highly refractive regions, reminiscent of the sky near a hot road in a summer day. Here, neighbouring light rays can end up in distinct regions of the sky, even distinct hemispheres, producing a chaotic mirage.

The second distinctive feature in Fig.~\ref{fig:Lensing_disk_a}, was already mentioned in the first scenario, but it is now clearer: the shadow varies from the familiar circular shape to a more prolate spheroid. In this scenario, the shadow is not blocked by the disk and this deformation can be fully appreciated. In Fig.~\ref{fig:shadow}  we exhibit just the BH shadow corresponding to the images in Fig.~\ref{fig:Lensing_disk_a}, to appreciate its dependence on $m$. One also observes, in Fig.~\ref{fig:Lensing_disk_a}, that the Einstein ring, which is perfectly circular for $m=0$, becomes oblate (as opposed to the prolate shadow), with some distortion near the equator.

\begin{figure}[h!]
\begin{center}
\includegraphics[width=0.3\textwidth]{mu40.jpg}\hspace{0.05cm}\includegraphics[width=0.3\textwidth]{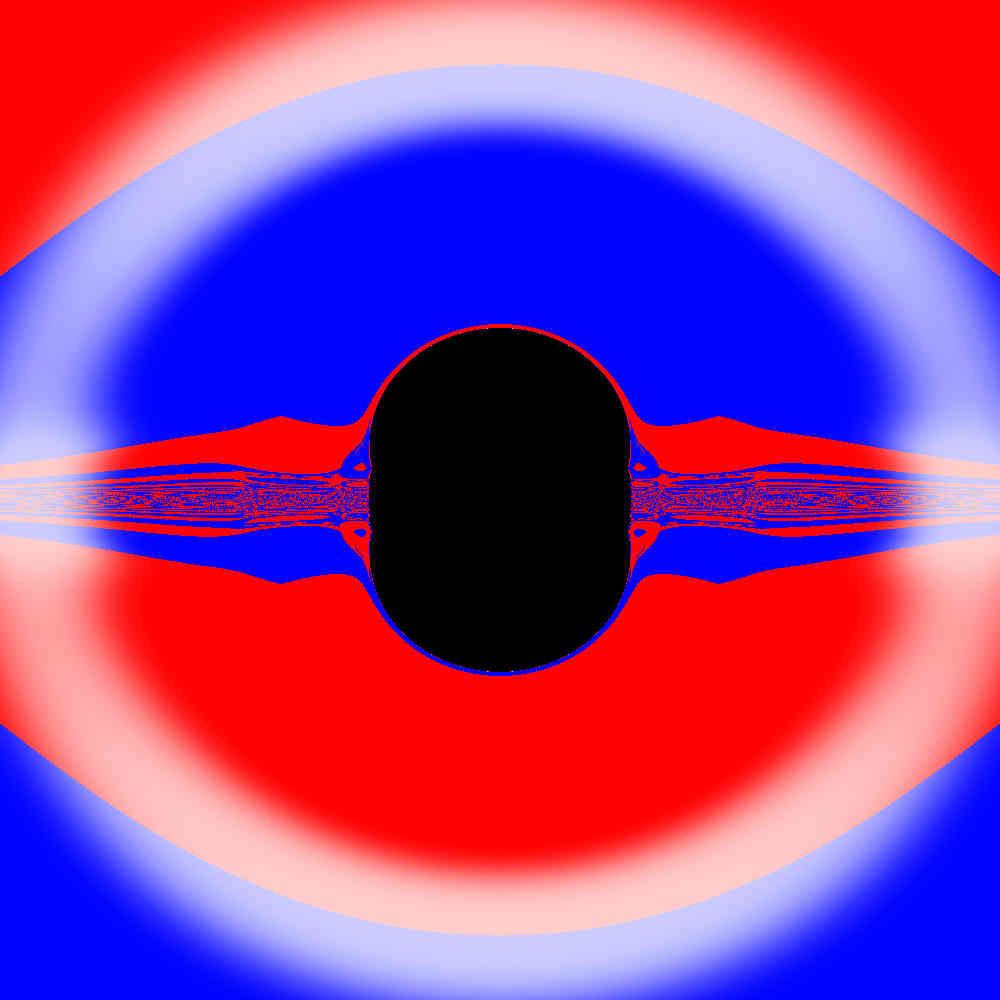}\\
\caption{\small A duplication of panel (f) of Fig.~\ref{fig:Lensing_disk_a} (left panel)  compared  with the same solution and setup but placing the celestial sphere at a larger radius, $\tilde{r}=400 $ (right panel).}
\label{fig:extra}
\end{center}
\end{figure}

As a closing remark, we have assessed how a different choice for the celestial sphere size impacts on the final results. In Fig.~\ref{fig:extra} we have recomputed panel (f) of Fig.~\ref{fig:Lensing_disk_a} with a much larger celestial sphere, almost $\simeq 13$ times larger. There are some minor changes in the lensing details, e.g. different sizes of the white ring and displacement of coloured regions. However, the main features are robust to this modification; in particular, the shadow is invariant under this operation.

\newpage

\section{Conclusions}
\label{section:Conclusions}
In this paper we have analysed the impact of having a heavy, backreacting, accretion disk around a BH in the latter's gravitational lensing and shadow. For this purpose we have used a Weyl solution describing the superposition of a Schwarzschild BH with a LL disk~\cite{Lemos_Letelier}, which has an inner edge. We chose the disk's inner edge and the fraction of the total spacetime mass in the disk to yield a physical configuration, where all disk particles have sub-luminal and real velocities and, moreover, all circular timelike geodesics in the disk region are stable. The inner edge of the disk was chosen to be at the ISCO, although this requirement could be relaxed.

The impact of the heavy disk in the images becomes more noticeably when the fraction of the total mass in the disk increases, as expected. The heavy disk bends light. This leaves a clear imprint on the trajectory of photons with a small $z$ component of their momentum that travel close to the disk. For an equatorial observer, and a disk lit scenario, this implies the thin disk appears thick. This thickening of the disk stretches the BH shadow, which becomes more prolate.

\begin{figure}[h!]
\begin{center}
\vspace{0.1cm}
\includegraphics[width=0.49\textwidth]{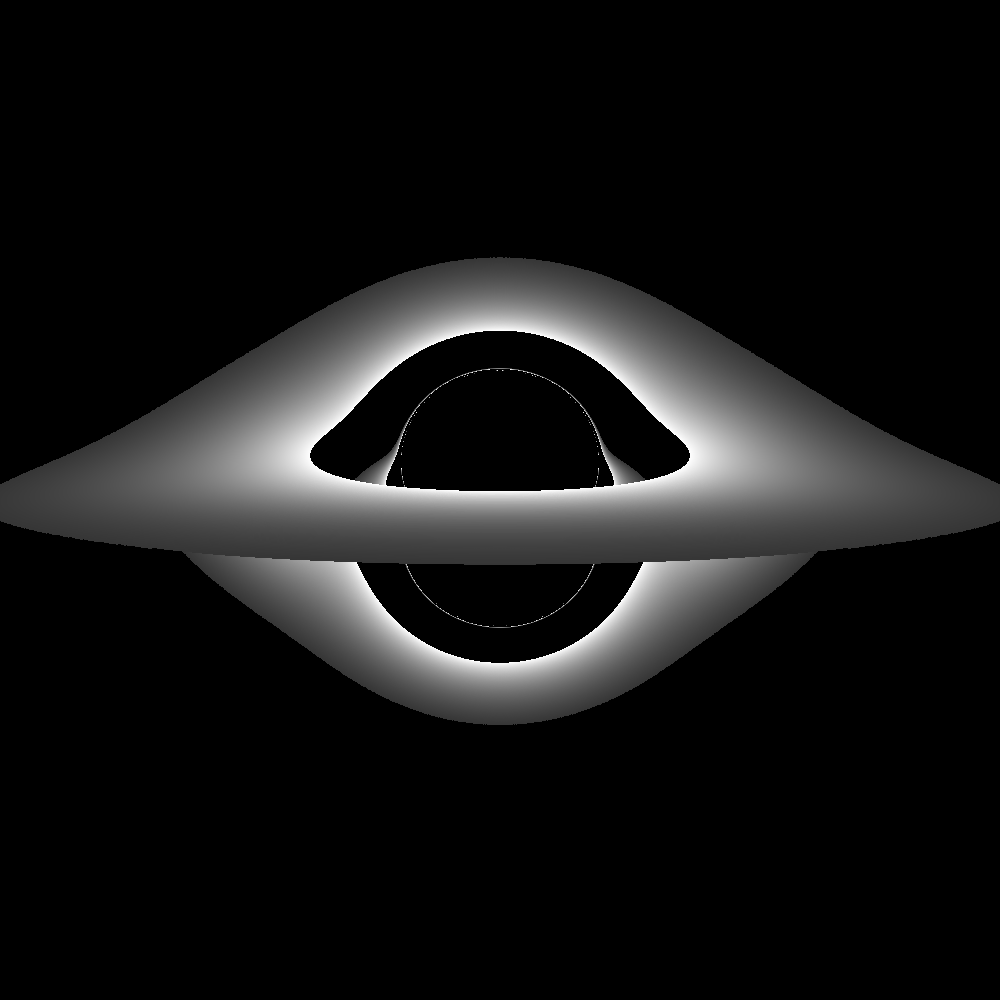}
\caption{\small A composed BH+LL disk system, corresponding to configuration (f) in Fig.~\ref{fig:AllowedRegions}. As in Fig.~\ref{fig0} the disk, which is the source of radiation, is on the equatorial plane ($\theta=90º$). It has an inner edge at the ISCO, a decaying luminosity profile, and an outer edge at certain radius. The observer is at $\theta=86º$ and at an areal radius of $20M$, where $M$ is the total mass of the system.
}
\label{figfinal}
\end{center}
\end{figure}

The disk used in Section~\ref{section:Shadow_case} was endowed with a homogeneous luminosity profile that extends to infinity. A more realistic visualisation can be obtained imposing a luminosity profile for the disk, which decays with increasing radius and terminates at some sufficiently large radius, providing the impression of a disk  outer edge. We recall  the LL disk extends all the way to spatial infinity. Under similar observation conditions as those used for Fig.~\ref{fig0}, which in particular considers the observer outside the equatorial plane, one obtains Fig.~\ref{figfinal}, using the configuration with highest $m$ analysed in this paper: solution (f) in Fig.~\ref{fig:AllowedRegions}. The most obvious differences with respect to Fig.~\ref{fig0} is the thickening of the disk and the smaller BH shadow, which gets stretched into an oblate shape.

For a scenario where the illumination comes from a ``far-away" celestial sphere and the disk is transparent, allowing photons to pass through, an observer close to the equatorial plane looking towards the sides of the shadow, will see a strong and chaotic refractive region in the vicinity of the equator. Again, photons with a small $z$ component of their momentum that travel close to the disk are bent towards the disk, but this time cross it and oscillate a few times around it  before escaping towards the celestial sphere, which occurs as they get to a larger radial coordinate and the disk becomes lighter. Then, open sets of observation angles lead to different numbers of oscillations, and to a different hemisphere endpoint of the photons in the far away celestial sphere. We have described this as a chaotic mirage. As expected, this feature becomes  weaker and more concentrated around the equator as one observers the disk further out (larger $\beta$ angles). This explains the spiky morphology of the strongly refractive region.

As a direction for further work, one could enlarge the analysis herein to also consider non-equatorial observations, as well as some illustrative examples in subregions $B$ and $C$. Moreover, it could be interesting to make the model more realistic by considering a spinning BH and/or a spinning disk.

\section*{Acknowledgements}
This work is supported by the Funda\c{c}\~ao para a Ci\^encia e a Tecnologia (FCT) project UID/MAT/04106/2019 (CIDMA), by CENTRA (FCT) strategic project UID/FIS/00099/2013, and by the project PTDC/FIS-OUT/28407/2017. This work has further been supported by  the  European  Union's  Horizon  2020  research  and  innovation  (RISE) programmes H2020-MSCA-RISE-2015 Grant No.~StronGrHEP-690904 and H2020-MSCA-RISE-2017 Grant No.~FunFiCO-777740. P. C. is supported by the Max Planck Gesellschaft through the Gravitation and Black Hole Theory Independent Research Group. We would like to thank M. Rodriguez for discussions. The authors would like to acknowledge networking support by the COST Action CA16104.  The authors thankfully acknowledge the computer resources, technical expertise and assistance provided by CENTRA/IST and by the University of Aveiro (UA). Computations were performed on the clusters ``Argus'' in the UA and ``Baltasar-Sete-Sóis'' in IST, the latter supported by the H2020 ERC Consolidator Grant ``Matter and strong field gravity: New frontiers in Einstein's theory" grant agreement no. MaGRaTh-646597.

\appendix
\section{Earlobes}
\label{section:appendix}
The right panel of Fig.~\ref{trajectory1} clearly shows two ``earlobes" on each side of the shadow, near the equatorial plane. These ``earlobes" can actually be seen in all three bottom panels of Fig.~\ref{fig:Shadow_disk_a}, becoming progressively more noticeable as $m$ increases. These features can also be seen in all three bottom panels of  Fig.~\ref{fig:Lensing_disk_a}. What is causing these ``earlobes"?

First, as a test on the robustness of this feature, we have checked how the size of the ``earlobes''  depends on the cutoff imposed by the celestial sphere size. Fig.~\ref{fig:extra} reveals that the earlobes' existence does not depend on this choice, albeit their size decreases when considering a further away celestial sphere. We have considered even more distant celestial spheres that corroborate this conclusion.

\begin{figure}[h!]
\begin{center}
\includegraphics[width=0.3\textwidth]{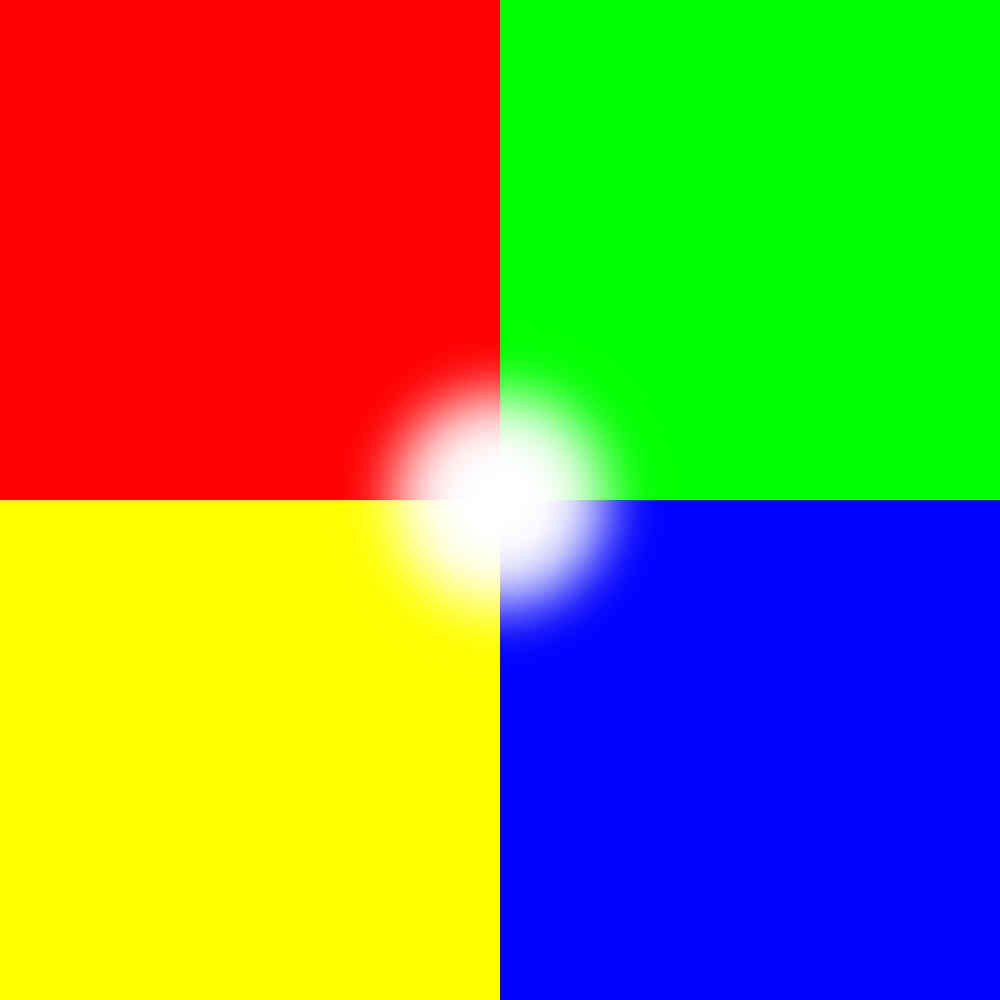}
\includegraphics[width=0.3\textwidth]{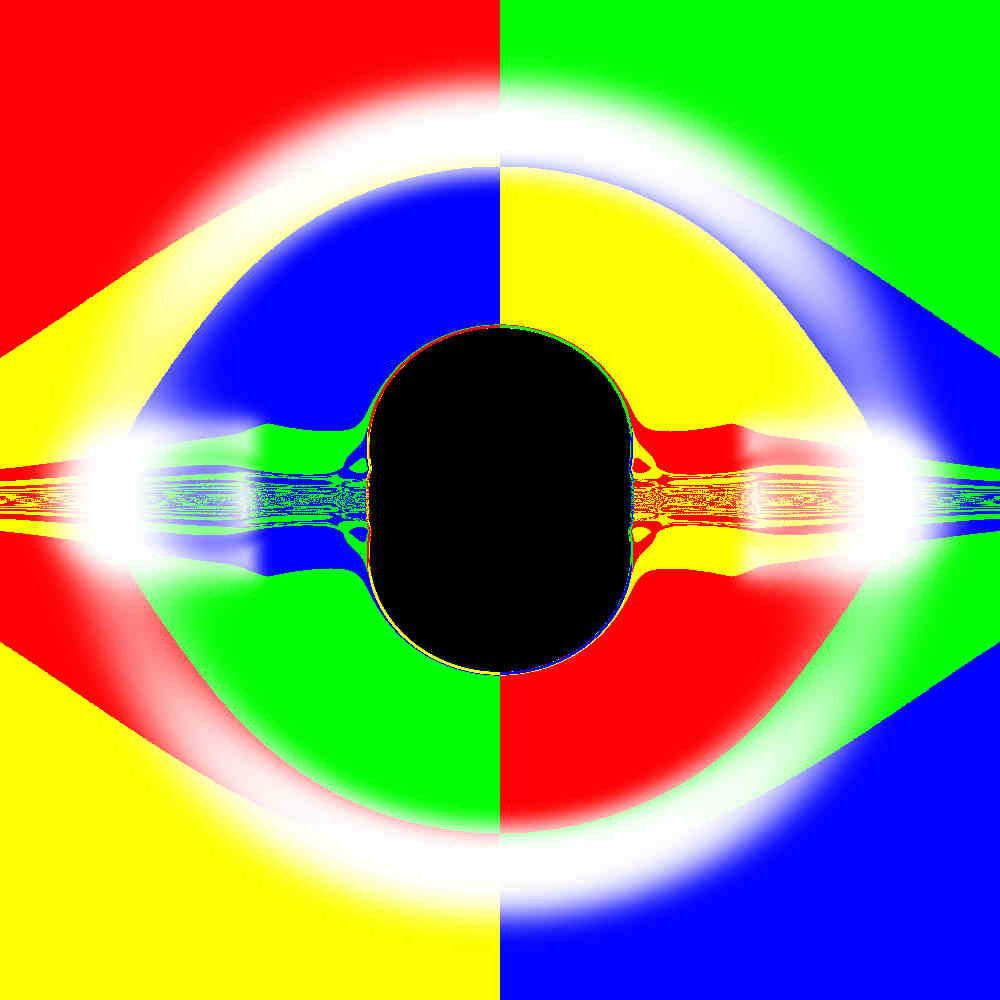}\hspace{0.05cm}
\includegraphics[width=0.3\textwidth]{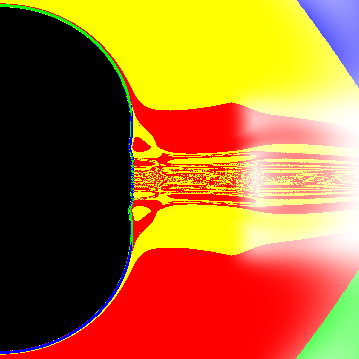}\hspace{0.05cm}
\caption{\small  (Left panel) Observation of an empty space with a celestial sphere with four colours. (Middle panel) Image of a BH+LL disk with $m=0.6$ with the same celestial sphere. (Right panel) Zoom of the middle panel, to better appreciate the structure of the ``earlobes".}
\label{fig:earlobes}
\end{center}
\end{figure}

Second, to get some insight into the origin of these ``earlobes'' we have considered a more distinct celestial sphere, with four colours, rather than the two used in Fig.~\ref{fig:Lensing_disk_a}. With respect to the observer's orientation these colours are: red (green) for the left (right) side of the North hemisphere, and yellow (blue) for the left (right) side of the South hemisphere. Such a division into four quadrants closely follows~\cite{Bohn2014}. In empty space (no BH or disk), this celestial sphere leads to the observation in Fig.~\ref{fig:earlobes} (left panel).

Placing now the configuration (f)  in  Fig.~\ref{fig:Shadow_disk_a} within this celestial sphere one obtains the image seen in Fig.~\ref{fig:earlobes} (middle panel).  With this setup it becomes clear that the top (bottom) earlobes on the right side of the image arise from trajectories that, approaching the BH on its right hand side (from the observer's perspective) bend left around the BH and end up on the left part of the North (South) hemisphere. Thus these ``earlobes"  are red (yellow) -  see zoom in Fig.~\ref{fig:earlobes} (right panel). If the trajectory has a larger impact parameter, the gravitational pull of the disk prevents this bending and the ``earlobe" terminates.  Fig.~\ref{fig:Shadow_disk_a},  moreover, shows that the earlobes correspond to photons that do not hit the disk. Further analysis of the corresponding trajectories shows that as these photons approach the BH they are pulled down by the disk's gravity but only cross the equatorial plane at $\rho<3$, thus beyond the inner edge. Then they are pulled back up by the disk on the opposite side, crossing again the equatorial plane before $\rho=3$, again avoiding hitting the disk. That is why the ``earlobes" are shown in grey in Fig.~\ref{fig:Shadow_disk_a}, thus corresponding to photons that hit numerical infinity rather than the disk.  Simultaneously, they are bent around the BH ending on the opposite meridional side of the celestial sphere.

\bibliography{biblio}{}
\bibliographystyle{JHEP}

\end{document}